\documentclass[fleqn,usenatbib]{mnras}

\usepackage{newtxtext,newtxmath}

\usepackage[T1]{fontenc}

\DeclareRobustCommand{\VAN}[3]{#2}
\let\VANthebibliography\thebibliography
\def\thebibliography{\DeclareRobustCommand{\VAN}[3]{##3}\VANthebibliography}

\usepackage{graphicx}	
\usepackage{amsmath}	
\usepackage{float}
\usepackage{orcidlink}  

\title[Environmental effects on galaxy evolution]{The influence of external environment at cosmic noon on the subsequent evolution of galaxy stellar mass}

\author[Gao et al.]{
Tianmu Gao,$^{\orcidlink{0000-0002-1158-6372} 1,2}$\thanks{E-mail: tianmu.gao@anu.edu.au}
J. Trevor Mendel,$^{\orcidlink{0000-0002-6327-9147}1,2}$
Lucas C. Kimmig,$^{\orcidlink{0009-0006-8337-8712}3}$
Claudia del P. Lagos,$^{\orcidlink{0000-0003-3021-8564}4,2}$
\newauthor Rhea-Silvia Remus,$^{\orcidlink{0009-0008-9260-7278}3}$
Emily Wisnioski,$^{\orcidlink{0000-0003-1657-7878}1,2}$
Kathryn Grasha$^{\orcidlink{0000-0002-3247-5321}1,2}$\thanks{ARC DECRA Fellow}
\\
$^{1}$Research School of Astronomy and Astrophysics, Australian National University, Weston Creek, ACT 2611, Australia\\
$^{2}$ARC Centre of Excellence for All Sky Astrophysics in 3 Dimensions (ASTRO 3D), Australia\\
$^{3}$Universit\"ats-Sternwarte, Fakult\"at f\"ur Physik, 
Ludwig-Maximilians-Universit\"at M\"unchen, Scheinerstr.\ 1, 81679 
M\"unchen, Germany\\
$^{4}$International Centre for Radio Astronomy Research, The University of Western Australia, 35 Stirling Highway, Crawley, WA 6009, Australia\\
}

\date{Accepted XXX. Received YYY; in original form ZZZ}

\pubyear{\the\year{}}

\begin{document}
\label{firstpage}
\pagerange{\pageref{firstpage}--\pageref{lastpage}}
\maketitle

\begin{abstract}
Connecting high-redshift galaxies to their low-redshift descendants is one of the most important and challenging tasks of galaxy evolution studies. In this work, we investigate whether incorporating high-redshift environmental factors improves the accuracy of matching high-redshift galaxies to their $z\sim0$ descendants, using data from the \textsc{Eagle} and \textsc{Magneticum} simulations. Using random forest regression, we evaluate the relative importance of a set of environmental metrics at $z\sim3$ in determining the stellar mass of descendant galaxies at $z\sim0$. We identify the spherical overdensity within 1 cMpc ($\delta_{1,\mathrm{sp}}$) as the most important environmental predictor. Tracking galaxies at $z\sim3$ with similar initial stellar masses but different $\delta_{1,\mathrm{sp}}$ values, we find that, across all mass bins in both simulations, high-density environments produce $z\sim0$ descendants with median stellar masses up to eight times higher than the descendants of galaxies in low-density environments. For galaxies with $M_{*}\lesssim10^{10}M_{\odot}$, the difference is attributable to more merger-induced mass growth in high-density environments, whereas for higher-mass galaxies, it results from a combination of enhanced in-situ star formation and greater external mass accretion. By assessing the importance of overdensity across multiple scales and redshifts, we find that at $z\gtrsim2$, environmental factors become as important as stellar mass in predicting the stellar mass of $z\sim0$ descendants. Compared to using stellar mass at $z\sim3$ alone, incorporating $\delta_{1,\mathrm{sp}}$ reduces the scatter in the residuals between the predicted and actual stellar masses by approximately 20\% in \textsc{Eagle} and 35\% in \textsc{Magneticum}.

\end{abstract}




\section{Introduction}\label{sec:intro}

One of the most important discoveries made by the \emph{Hubble Space Telescope (HST)} is the peak of cosmic star formation rate density at $z\approx2$, approximately 10 billion years ago, an epoch also known as the 'cosmic noon' \citep{1996Lilly,1996Madau,1998Madau}. After this epoch, the most massive galaxies began to quench, while the average star formation rate (SFR) of star-forming galaxies also significantly declined. Meanwhile, over this period, galaxies also undergo drastic morphological transformations, evolving from clumpy disks at high redshift \citep[e.g.,][]{2009Schreiber,2015Wisnioski} into the diverse morphological types observed today -- a process often referred to as the emergence of the Hubble sequence. It has been shown that several physical processes occurring within galaxies may contribute to the above evolution, including stellar feedback \citep[e.g.,][]{1990Heckman}, active galactic nuclei (AGN) feedback \citep[e.g.,][]{2012Fabian}, reduced gas accretion \citep[e.g.,][]{2020Walter}, disk instability \citep[e.g.,][]{2015Zolotov}, and morphological quenching \citep[e.g.,][]{2009Martig}. On the other hand, mechanisms related to the interactions between galaxies and their external environment, such as galaxy mergers, ram-pressure stripping, and cosmological starvation, could also contribute to the changes in galaxy properties \citep[e.g.,][]{2018Lagos,2021Cortese,2015Peng}.

All these competing mechanisms, each with substantially different timescales of influence and varying dependencies on stellar mass and environment, often simultaneously influence galaxy evolution, leading to different formation pathways \citep[e.g.,][]{2022Lagos,2025Lagos,2025Kimmig}. Therefore, disentangling the influence of them requires tracing the evolution of galaxies with different stellar masses and external environments over a wide range of cosmic time (\citealp[e.g., see the review of][]{2017Naab}).

With the increased sensitivity enabled by \emph{JWST}, space-based spectroscopic observations are now capable of detecting faint emission lines from ionised gas, as well as the absorption line continuum from galaxies at cosmic noon, allowing their stellar populations and chemical compositions to be constrained in much greater detail. For example, \emph{JWST}/NIRSpec \citep{2022Jakobsen} $R\sim1000$ slit spectra of cosmic noon galaxies have revealed the prevalence of AGN-driven multi-phase gas outflows \citep{2024Davies,2024Belli,2024Bugiani}, whose decisive role in the rapid quenching of massive quiescent galaxies is further supported by the reconstruction of the star formation histories of these galaxies \citep{2024Park}. These new measurements have prompted further efforts to understand evolutionary pathways and compare galaxy populations over cosmic time in the \emph{JWST} era.

To fully understand the subsequent evolution of cosmic-noon galaxies, such as the quenching of star formation and morphological transformations, we need to track the continuous evolution of different galaxy populations over the past 10 billion years. Since we cannot directly observe the long-term evolution of an individual galaxy, a key challenge is to construct physically motivated evolutionary links between galaxies observed at different cosmic epochs. A common approach to connecting galaxies across different redshifts is to use a constant cumulative co-moving number density (hereinafter referred to as 'number density') (\citealp[e.g.,][]{2010vanDokkum,2013vanDokkum,2011Papovich,2015Papovich,2011Bezanson,2011Brammer,2013Patel,2013Muzzin,2014Ownsworth,2016Ownsworth,2015Shankar,2015Huertas-Company}). In this approach, a galaxy's number density is defined as the number of galaxies per unit volume with a stellar mass equal to or greater than that of the targeted galaxy. This method accounts for the overall mass evolution of a galaxy population and can be easily applied when the cumulative mass function (CMF) is established at all redshifts of interest. By determining the number density of a galaxy population with similar stellar masses at one redshift and identifying the corresponding stellar masses at other redshifts, one can establish links between galaxies across different epochs. 

Establishing progenitor-descendant links at a constant number density assumes that (1) the total number of galaxies and (2) their rank order remain unchanged over time. However, these assumptions may not hold due to (1) galaxy mergers, which change the total number of galaxies, and (2) the stochastic growth, including in-situ star formation and ex-situ accretion from mergers. As a result, the median mass evolution tracks of galaxy populations with different stellar masses—derived from constant number density selection—are offset by a factor of a few from those measured directly from galaxy merger trees in hydrodynamical simulations (\citealp[e.g.,][]{2015Torrey}). Alternatively, the number density evolution has been analysed in abundance matching models \citep{2013Behroozi}, semi-analytic models \citep{2013Leja} and hydrodynamical simulations \citep{2015Torrey,2017Torrey,2016Jaacks,2016Clauwens}. These various approaches have consistently shown similar trends in terms of the number density evolution, even when galaxy ranks are assigned based on dark matter halo mass or stellar velocity dispersion instead of stellar mass. This consistency suggests that the number density evolution is more fundamental and more closely linked to the $\mathrm{\Lambda}$CDM framework compared to the evolution of galaxies' intrinsic properties. Progenitor-descendant matching based on evolving number density selection has successfully reproduced the median mass evolution of populations with different stellar masses, as defined by merger trees in simulations \citep{2015Torrey} and has been applied to observational datasets (\citealp[e.g.,][]{2014Marchesini,2015Salmon,2015Papovich}). Nevertheless, the inferred $z\sim0$ descendants of high-redshift galaxies with similar masses still exhibit a $\sim$1 dex scatter in stellar mass (\citealp[e.g.,][]{2015Torrey}), which significantly limits the accuracy of any inferred continuous evolutionary tracks.

Cosmological simulations have suggested a 'two-phase' model for the formation of massive galaxies (\citealp[e.g.,][]{2010Oser}). In this model, a rapid early phase occurs at $z>2$, during which 'in-situ' stars form within galaxies from cold gas. This is followed by an extended phase since $z\simeq3$, during which 'ex-situ' stars are primarily accreted from surrounding satellite galaxies that formed at $z>3$. Building on this, we expect that by incorporating the surrounding environment as a proxy for ex-situ accretion, we can better characterise the evolutionary pathways of stellar mass in galaxies across different environments, thereby facilitating a more accurate matching of high-redshift galaxies to their low-redshift descendants. To examine the extent to which the external environment influences the stellar mass growth of galaxies, we investigate the late-time stellar mass evolution of simulated galaxies across different high-redshift environments using data from the \textsc{Eagle} \citep{2015Crain,2015Schaye,2016McAlpine} and the \textsc{Magneticum} simulations \citep{2015Teklu,2016Dolag,2025Dolag}. Leveraging their large simulated volumes and merger trees, we can quantify environments across different scales while tracking the evolution of galaxy properties over cosmic time. 

The remainder of this paper is structured as follows. In Section \ref{sec:method}, we introduce the simulations used in this study and describe how we quantify the external environment of galaxies. We then use random forest regression, a machine learning method, to quantify the relative importance of environments at different scales, as measured by various environmental metrics, in determining the stellar mass of the local descendants of high-redshift galaxies in Section \ref{sec:random forest}. Based on the results of the random forest analysis, we select key metrics and analyse the evolution of number density and stellar mass of galaxies across different high-redshift environments, as quantified using these metrics, in Section \ref{sec:results}. We further investigate the reasons why the environment affects the subsequent stellar mass growth of high-redshift galaxies with different stellar masses and discuss the implications of our findings in Section \ref{sec:discussion}. Finally, we summarise our main findings in Section \ref{sec:conclusion}.

\section{Methods}\label{sec:method}

In this study, we use galaxy catalogues from the \textsc{Eagle} simulation and the \textsc{Magneticum} simulation. By using multiple simulations, we can ensure that any general findings consistent across them are genuine and not merely the result of specific physical models in one simulation. In this section, we first briefly introduce each simulation, focusing on the physical models they employ related to stellar mass accumulation. Specifically, we highlight the halo merger trees along with the subgrid physics used to model in-situ star formation and feedback processes. We then introduce a series of environmental metrics from the literature, which are used to characterise the environments of simulated galaxies at different scales.

\subsection{\textsc{Eagle}}\label{sec:Eagle}

The 'Evolution and Assembly of GaLaxies and their Environment' (\citealp[\textsc{Eagle};][]{2015Crain,2015Schaye,2016McAlpine}) project consists of a series of hydrodynamic simulations based on a modified version of the \textsc{Gadget}-3 code \citep{2005Springel}, incorporating the smoothed particle hydrodynamics (SPH) improvements described in \cite{2015Schaller}, with varying box size, particle resolution and subgrid physics. It adopts a flat $\mathrm{\Lambda}$CDM Universe with $\Omega_{\mathrm{m}}= 0.307$, $\Omega_{\mathrm{\Lambda}}= 0.693$ and $h\equiv H_{0}/(100\,\rm km\,s^{-1}\,Mpc^{-1})=0.6777$ \citep{2014Planck}. Based on our need to study the external environments of galaxies at various scales, in this study we select the largest box from the suite, Ref-L0100N1504, which has a simulated volume of $(67.8\,\mathrm{cMpc}/h^{-1})^{3}$, an initial baryonic particle mass of $1.81\times10^{6}M_{\odot}$, a dark matter particle mass of $9.7\times10^{6}M_{\odot}$, and includes an initial number of $2\times1504^{3}$ particles (an equal number of baryonic and dark matter particles).

In \textsc{Eagle}, dark matter halos are first identified using the friends-of-friends (FoF) method \citep{1985Davis}. Baryonic particles are then assigned to the FoF-halo that contains their nearest dark matter neighbour. Finally, galaxies are defined as gravitationally bound subhalos residing within each halo, as identified by the \textsc{Subfind} algorithm \citep{2001Springel,2009Dolag}. A detailed description of the \textsc{Eagle} merger trees, which explains how individual galaxies are tracked across simulation snapshots, can be found in \cite{2017Qu}.

A metallicity- and temperature-dependent local density threshold is used to determine which gas particles are forming stars \citep{2004Schaye}. Star-forming gas is stochastically converted into stars according to a reformulated Kennicutt–Schmidt star formation law, where the star formation rate scales with local gas pressure \citep{2008Schaye,2015Schaye}. Each star particle models a simple stellar population with a \cite{2003Chabrier} initial mass function (IMF) in the range 0.1--100\,$M_{\odot}$. Stellar feedback triggered by a single stellar population is implemented by injecting thermal energy into random surrounding gas particles near young star particles. Energy injection depends on the local gas density and metallicity \citep{2015Crain}, and alway instantly heats the gas particles by a temperature increment of $\Delta T=10^{7.5}\mathrm{K}$ \citep{2012Dalla}. AGN feedback is implemented in a manner similar to stellar feedback. Thermal energy is injected into the ISM surrounding the accreting supermassive black hole (SMBH) with a constant average efficiency related to the accretion rate, raising the temperature of the ISM by $\Delta T=10^{8.5}\mathrm{K}$ \citep{2009Booth}. The free parameters in the subgrid model for feedback are calibrated to reproduce the galaxy stellar mass functions (GSMF) \citep{2009Li,2012Baldry}, mass–size relation \citep{2003Shen,2012Baldry} and $M_{\mathrm{BH}}$--$M_{*}$ relation \citep{2013McConnell} at $z\approx0$.

In this study, we use snapshots numbered 12 to 28, corresponding to redshifts from 3.02 to 0. For the diffuse particles at large radii that are still bound to the corresponding subhalo, we use aperture masses and star formation rates of galaxies within spherical apertures of 30 pkpc in this study, as recommended by \cite{2016McAlpine}.

\subsection{\textsc{Magneticum}}\label{sec:Magneticum}

The \textsc{Magneticum} simulations \citep{2015Teklu,2016Dolag,2025Dolag} are a suite of simulations performed using an updated version of the \textsc{Gadget-2} code, which includes improved implementations of the SPH \citep{2004Dolag,2005Dolag}. All simulations were conducted using the WMAP7 $\mathrm{\Lambda}$CDM cosmology with Universe with $\Omega_{\mathrm{m}}= 0.272$, $\Omega_{\mathrm{\Lambda}}= 0.728$ and $h\equiv H_{0}/(100\,\rm km\,s^{-1}\,Mpc^{-1})=0.704$ \citep{2011Komatsu}. In this work, we use the Box4-uhr run, with a box size of $68^{3}\,\mathrm{cMpc}^{3}$ and an initial particle number of $2\times576^{3}$ (dark matter and gas). The mass resolution for dark matter and gas particles is $m_{\mathrm{DM}}=5.1\times10^{7}M_{\odot}$ and $m_{\mathrm{gas}}=1.0\times10^{7}M_{\odot}$, respectively. Since each gas particle can convert into up to four stellar particles, this results in an average stellar particle mass of $m_{*}\approx\ 1/4 m_{\mathrm{gas}}\approx2.6\times10^{6}M_{\odot}$. 

The identification of dark matter halos and galaxies is performed using a modified version of \textsc{Subfind} \citep{2001Springel,2009Dolag}. The evolution of galaxies through time is traced using \textsc{L-BaseTree} \citep{2005Springel}, allowing us to track the stellar mass assembly process of galaxies from $z=3$ to $z=0$.

The modeling of star formation and stellar feedback processes follows the prescription by \cite{2003Springel}. Each stellar particle represents a stellar population with a Chabrier IMF \citep{2003Chabrier}. In this star formation model, the ISM is considered to be two-phase, where the hot gas is heated by supernovae and can evaporate the cold clouds. It is assumed that 10\% of massive stars end their lives as Type II supernovae (SNe II). The galactic wind triggered by SNe II has a mass loading rate proportional to the SFR, and a wind velocity of $v_{\mathrm{wind}}=350\,\mathrm{km\,s^{-1}}$. The subgrid physics of AGN feedback is implemented via the model described by \cite{2014Hirschmann}, which is based on the framework from \cite{2010Fabjan} and incorporates a transition from quasar-mode to radio-mode AGN feedback. Both modes are implemented thermally, with the transition triggered when the SMBH accretion rate falls below an Eddington ratio of $f_{\mathrm{edd}}<10^{-2}$ (see also \citealp{2007Sijacki}). The radio mode is assigned a feedback efficiency four times higher than that of the quasar mode, to account for the heating of the surrounding ICM through AGN jets originating from those slowly-accreting SMBHs. The \textsc{Magneticum} simulations are calibrated to match the intracluster gas content of observed galaxy clusters, rather than to reproduce specific properties of individual galaxies \citep[e.g.,][]{2024Popesso}.

All 104 snapshots from $z\approx3$ to $z\approx0$ are used in this work to achieve a redshift range similar to that of \textsc{Eagle}, allowing for the direct comparison of galaxy evolutionary tracks between these two simulations.

\subsection{Environmental Measures}\label{sec:metrics}

There are various metrics from the literature that can be used to measure the external environment of galaxies at different scales. In this work, we quantify galaxy environments using a series of measures analysed in \cite{2012Muldrew}, which can be categorised into two types: nearest neighbour measures and fixed-aperture measures.

The nearest neighbour projected surface density can be defined as:

\begin{equation}
\Sigma_{n}=\log\frac{n}{\pi\,r^{2}_{n}},
\end{equation}

\noindent
where $n$ is the number of neighbouring galaxies within the projected comoving distance to the $n$-th nearest neighbour $r_{n}$, in the $x$–$y$ plane. Instead of using the 3D distance, which is not feasible in real observations, we adopt a velocity cut of $\pm1000\,\mathrm{km\,s^{-1}}$ to account for the separation in the third dimension, i.e., along the redshift $z$-axis. It should be noted that the velocity difference between two galaxies along the line of sight is the combined result of their peculiar velocities and the Hubble expansion. The line of sight velocity of galaxy $i$ with respect to galaxy $j$ can be expressed as:

\begin{equation}
    \Delta v_{\rm LOS}=v_{z,i}-v_{z,j}+\frac{H(z)}{1+z}d_{z},
\end{equation}

\noindent
where $v_{z,i}$ and $v_{z,j}$ are the peculiar line-of-sight velocities of galaxies $i$ and $j$, $d_{z}$ is their comoving separation along the $z$-axis, and $H(z)$ is the Hubble parameter at the corresponding redshift. We choose the number of nearest neighbours to
be $n= 3, 7, 10, 20\ \rm{and}\ 32$, and also include the average density of the 4th and 5th nearest neighbours, as used by \cite{2006Baldry}.

Fixed-aperture measures are often expressed in terms of overdensity. In this work, we first calculate the number of neighbouring galaxies $N_{g}$ within a fixed-size 2D aperture of radius $r$, centered on each galaxy in the simulation box, and then normalise each value by the mean neighbour count $\bar{N_{g}}$ calculated across all galaxies. Thus, the overdensity $\delta_{r}$ is defined as:

\begin{equation}
\delta_{r}=\frac{N_{g}-\bar{N_{g}}}{\bar{N_{g}}}.
\end{equation}

\noindent
In this paper, we adopt $1, 2, 5\ {\rm{and}}\ 10$ cMpc as the radii of the apertures (hereafter $\delta_{r,\mathrm{ap}}$), and also include the annulus measures listed in Table 1 of \cite{2012Muldrew} (hereafter $\delta_{r,\mathrm{an}}$), which were first proposed in \cite{2010Wilman}. Similar to the calculation of nearest neighbour densities, we only include neighbouring galaxies with line-of-sight velocities within $\pm1000\,\mathrm{km\,s^{-1}}$ of the target galaxies. Additionally, assuming the 3D distances are known, we also include overdensities within spherical regions of radii $4, 8\ {\rm{and}}\ 12$ cMpc (hereafter $\delta_{r,\mathrm{sp}}$) in our analysis, as originally adopted in \cite{2012Muldrew}, along with an additional radius of $1$ cMpc explored in this study. In total, we include 20 different environmental metrics in this study.

Taking into account the resolution of each simulation, we only include galaxies with stellar masses greater than $10^{9}\,M_{\odot}$. This ensures that all galaxies included in our analysis have $\gtrsim500$ stellar particles.

\section{Random Forest Analysis}\label{sec:random forest}

The goal of this section is to evaluate how useful each environmental metric is, compared to other metrics, in driving stellar mass growth of galaxies. To this end, we employ a random forest regression analysis approach to predict the stellar mass of the $z=0$ descendants of galaxies with stellar masses greater than $10^{9}\,M_{\odot}$ in the $z\sim3$ snapshot across both simulations. We first introduce our methodology, followed by the detailed results for each simulation.

\subsection{Method}\label{sec:random forest description}

Random Forest (RF; \citealt{2001Breiman}) is a supervised machine learning algorithm that combines the outputs of multiple decision trees to perform classification or regression tasks. A decision tree begins with a root node that contains the entire training dataset, whose samples include both input features and target variables, and then splits into branches according to conditions applied to input features. Each split aims to minimise the variability of the target variable within the resulting nodes, which in turn reduces the expected prediction error and allows the model to capture the dependence of the target variable on the input features. This process continues until a stopping criterion is satisfied, such as a maximum depth of the tree or a minimum number of samples per node. The final output of a leaf is typically the average of the target variable values of the samples it contains. For model evaluation in regression tasks, where the target variable is continuous, the prediction error is often measured using metrics such as mean squared error (MSE). 

In a decision tree, feature importance reflects how much each feature contributes to reducing the prediction error during the splitting process. In regression tasks, this reduction is quantified as the decrease in MSE when a node is split based on a particular feature. These reductions are then summed across all nodes in the tree where the feature is used, yielding an overall importance score.

A random forest is created by aggregating the predictions of numerous decision trees, each trained on a random subset of the data and using a random subset of features. This combination improves the predictive performance of the model and reduces the risk of overfitting. The final prediction of a regression task is obtained by averaging the predictions of all trees in the forest. Similarly, overall feature importance is calculated by averaging the individual importance scores from each tree. These scores are then normalised to provide a relative measure of each feature's contribution to the predictive performance of the model.

\subsection{Implementation}\label{sec:random forest parameters}

For the RF analysis, we use the \textsc{RandomForestRegressor} from the \textsc{Scikit-Learn} python package \footnote{https://scikit-learn.org/stable/index.html} \citep{scikit-learn}. For \textsc{Eagle} and \textsc{Magneticum}, the input data samples consist of 3860 and 5603 galaxies, respectively. The external environments of all these galaxies, quantified using 20 different metrics mentioned above, along with their stellar masses, are used as features for the regression. The stellar mass of their $z=0$ descendants, derived from merger trees, serves as the target prediction. In addition to training the model using the entire dataset, we also split the galaxies into three mass bins based on their stellar masses at $z\sim3$: $\log (M_{*}/M_{\odot}) \leq 9.5$, $9.5 < \log (M_{*}/M_{\odot}) \leq 10$, and $\log (M_{*}/M_{\odot}) > 10$. In \textsc{Eagle}, these bins contain 2338, 962, and 560 galaxies, respectively, while in \textsc{Magneticum}, the bins contain 3854, 988, and 761 galaxies.

For each mass bin, we used a different parameter grid for parameter tuning. This approach ensures a balance between model capacity and generalisation ability, reducing the risk of both underfitting and overfitting. The parameter grid includes five parameters: the number of decision trees \texttt{n\_estimators}, the maximum depth of individual decision trees \texttt{max\_depth}, the minimum number of samples needed for a split \texttt{min\_samples\_split}, the minimum number of samples per leaf \texttt{min\_samples\_leaf} and the number of features considered when splitting each node \texttt{max\_features}. The \texttt{n\_estimators} parameter controls the total number of trees in the forest, and increasing it generally improves the model performance. However, more trees lead to higher computational costs. In practice, we find that increasing the number of trees exceeding 100 yields negligible changes in both the model performance and the resulting relative importance (hereafter, R.I.) values. Limiting the depth of the trees (\texttt{max\_depth}) helps prevent overfitting by controlling model complexity, as deeper trees can capture more complex relationships but also increase the risk of overfitting the data. Similarly, adjusting the \texttt{min\_samples\_split} and \texttt{min\_samples\_leaf} enables fine-tuning of the tree structure, with higher values producing simpler trees that can enhance generalisation ability but may increase bias by oversimplifying the model. Finally, the number of features considered at each split (\texttt{max\_features}) modulates the randomness of the splits: lower values increase diversity among trees and mitigate overfitting, while higher values allow finer splits at the cost of increasing risk of overfitting.

\begin{figure*}

\centering
    \includegraphics[width=\textwidth]{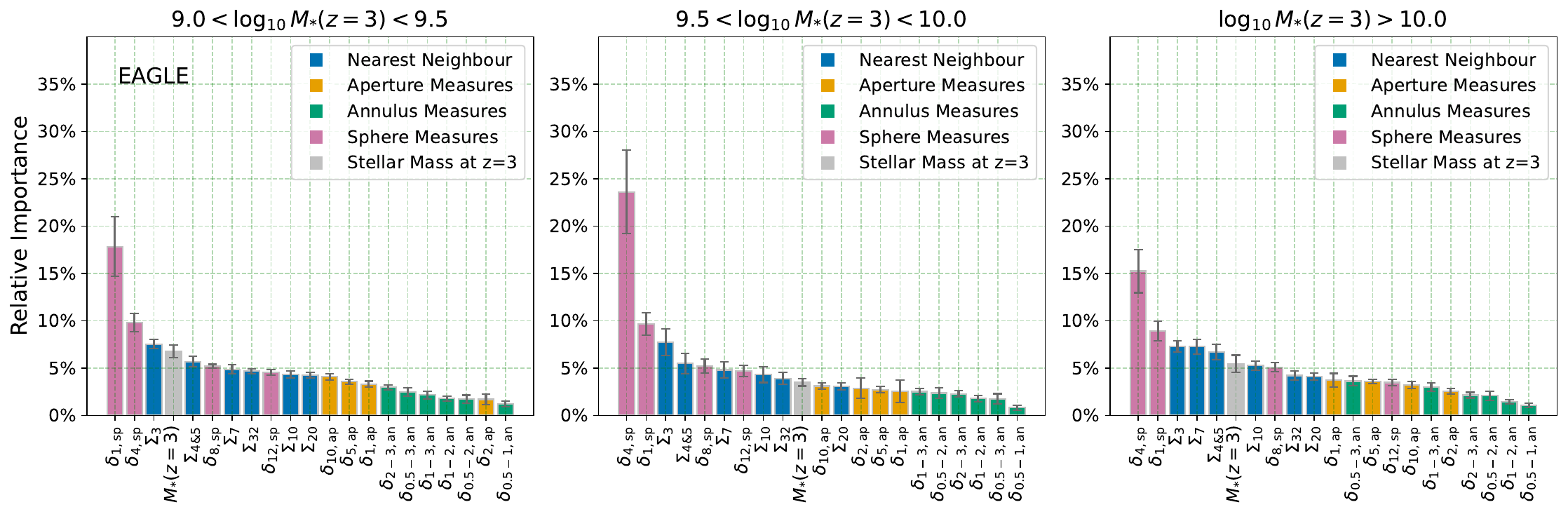}
    \includegraphics[width=\textwidth]{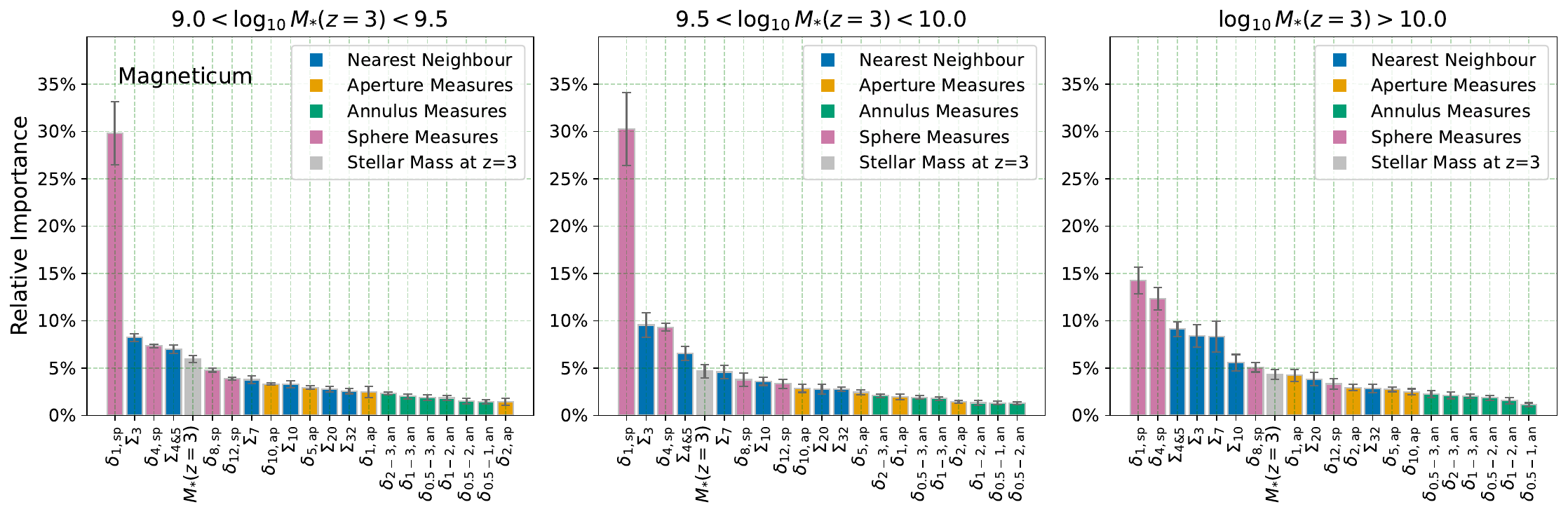}
    \caption{The relative importance (R.I.) of both the $z\sim3$ environment, as characterised by different metrics, and the stellar mass of the progenitor galaxies themselves, in determining the stellar mass of the $z\sim0$ descendants of galaxies with $9.0<\log (M_{*}/M_{\odot})\leq9.5$ (left), $9.5<\log (M_{*}/M_{\odot})\leq10.0$ (middle), and $\log (M_{*}/M_{\odot})>10.0$ (right) in \textsc{Eagle} (top) and \textsc{Magneticum} (bottom). The height of each bar denotes the R.I. of each feature, with the error bars representing the standard deviation from 10 independent runs. The bars are colour-coded by feature categories: nearest neighbour measures (blue), aperture measures (yellow), annulus measures (green), sphere measures (pink), and the stellar mass of the progenitor galaxies (gray).}
    \label{fig:rfr mass bins}
    
\end{figure*}

Each training run evaluates all parameter combinations in the corresponding parameter grid and selects the best model based on the negative MSE (\texttt{neg\_mean\_squared\_error}), which computes the MSE between predicted and true values on the validation set, and returns its negative so that higher scores indicate better model performance. Once the optimal parameter combination is determined, the model is trained on the full training set again and evaluated on a separate test set. We assess the model performance by comparing the root MSE (RMSE) on the training and test sets, and consider a run as validated only if the difference between them is less than 0.5. For each feature, we finally take the median importance across 10 validated runs as the final importance value, with the standard deviation representing the uncertainty.

\subsection{Results}\label{sec:random forest results}
In Figure \ref{fig:rfr mass bins}, we present the results of the RF analysis on the R.I. of external environment at different scales of progenitor galaxies in predicting the stellar mass of descendant galaxies, with the left, middle, and right panels corresponding to stellar mass ranges of $9.0<\log (M_{*}/M_{\odot})\leq9.5$, $9.5<\log (M_{*}/M_{\odot})\leq10.0$, and $\log (M_{*}/M_{\odot})>10.0$, respectively. Results from \textsc{Eagle} are shown in the top panel, while results from \textsc{Magneticum} are shown in the bottom panel. The height of each bar denotes the R.I. of each feature in terms of its ability for predicting the stellar mass of the descendants, with the error bars representing the standard deviation from 10 independent runs. All features are listed on the x-axis in order of importance, from highest to lowest. We also colour code the bars by feature categories: nearest neighbour measures (blue), aperture measures (yellow), annulus measures (green), sphere measures (pink), and the stellar mass of the progenitor galaxies (gray).

In \textsc{Eagle}, the spherical overdensities within radii of 1.0 and 4.0 cMpc ($\delta_{1,\mathrm{sp}}$ and $\delta_{4,\mathrm{sp}}$) are the two most important predictors across all three stellar mass bins. For the three mass bins, the importance of $\delta_{1,\mathrm{sp}}$ is $\mathrm{R.I.}=17.8\pm3.2\%\,(9.0<\log (M_{*}/M_{\odot})\leq9.5)$, $9.7\pm1.2\%\,(9.5<\log (M_{*}/M_{\odot})\leq10.0)$ and $8.9\pm1.0\%\,(\log (M_{*}/M_{\odot})>10.0)$, respectively; while the importance of $\delta_{4,\mathrm{sp}}$ is $\mathrm{R.I.}=9.8\pm1.0\%\,(9.0<\log (M_{*}/M_{\odot})\leq9.5)$, $23.6\pm4.4\%\,(9.5<\log (M_{*}/M_{\odot})\leq10.0)$ and $15.2\pm2.3\%\,(\log (M_{*}/M_{\odot})>10.0)$, respectively. Moreover, across all three stellar mass bins, the mean nearest neighbour density for $n=3$ ($\Sigma_3$) consistently ranks as the third most important metric, with its importance being $\mathrm{R.I.}=7.5\pm0.5\%\,(9.0<\log (M_{*}/M_{\odot})\leq9.5)$, $7.7\pm1.4\%\,(9.5<\log (M_{*}/M_{\odot})\leq10.0)$ and $7.3\pm0.6\%\,(\log (M_{*}/M_{\odot})>10.0)$, respectively. Overall, it is apparent that nearest neighbour densities with smaller $n$ and sphere overdensities with smaller radii have higher R.I. compared to those with larger $n$ and radii. In comparison, aperture/annulus overdensities generally have lower R.I. ($<5\%$).

In the model trained with \textsc{Magneticum} data, despite the implementation of many different subgrid physics models in two simulations, $\delta_{1,\mathrm{sp}}$, $\delta_{4,\mathrm{sp}}$ and $\Sigma_3$ remain the most predictive variables. $\delta_{1,\mathrm{sp}}$ ranks as the most important environmental metric throughout all stellar mass bins, with its R.I. being $29.8\pm3.3\%,(9.0<\log (M_{*}/M_{\odot})\leq9.5)$, $30.3\pm3.8\%\,(9.5<\log (M_{*}/M_{\odot})\leq10.0)$, and $14.3\pm1.4\%\,(\log (M_{*}/M_{\odot})>10.0)$, respectively. Moreover, aperture/annulus overdensities are generally less important. This suggests that, despite the different subgrid physics implementations and what they are tuned to, both cosmological simulations show that a similar subset of environmental tracers is most predictive for the $z=0$ stellar mass.

\begin{figure}
    \centering
    \includegraphics[width=\linewidth]{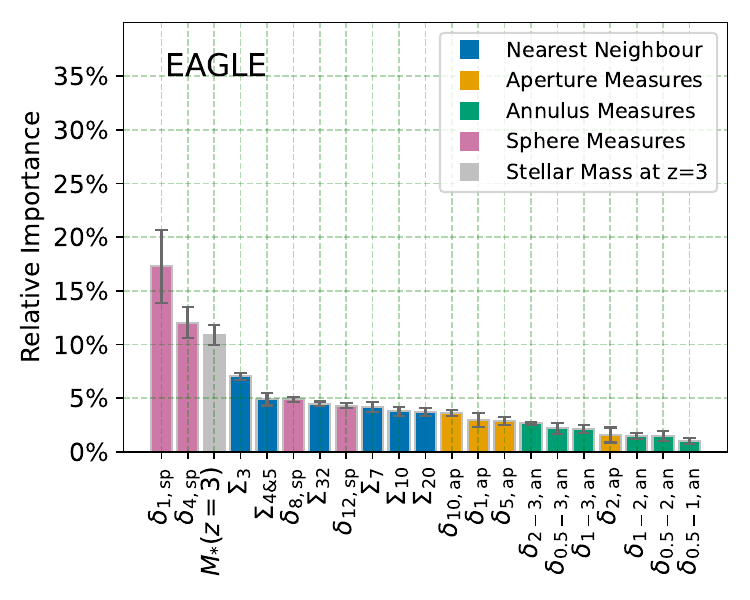}
    \includegraphics[width=\linewidth]{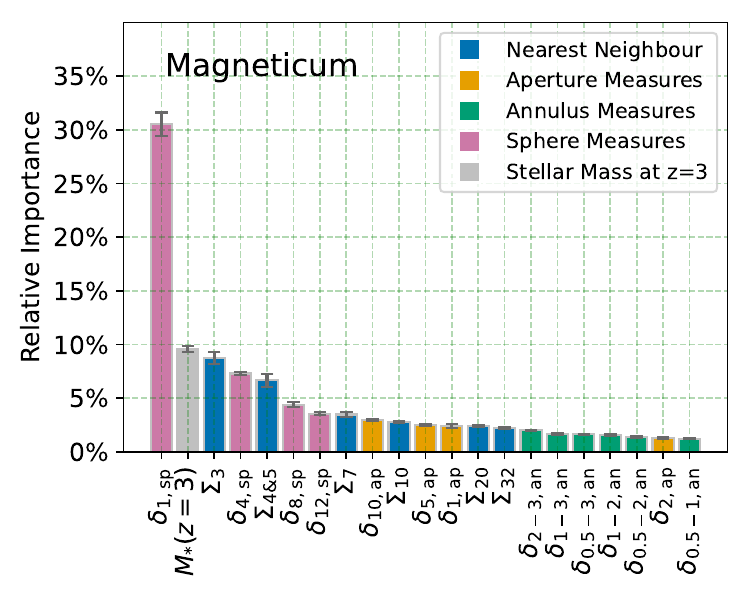}
    \caption{Same as Figure \ref{fig:rfr mass bins}, but for all galaxies with $\log (M_{*}/M_{\odot})>9.0$ in \textsc{Eagle} (top) and \textsc{Magneticum} (bottom).}
    \label{fig:rfr all mass}
\end{figure}

We also train models using all galaxies with $\log (M_{*}/M_{\odot})>9.0$ in \textsc{Eagle} and \textsc{Magneticum}, without dividing them into separate mass bins. The results are shown in Figure \ref{fig:rfr all mass}, with \textsc{Eagle} in the top panel and \textsc{Magneticum} in the bottom panel. In \textsc{Eagle}, $\delta_{1,\mathrm{sp}}$ has become the most important environmental feature, while the overall ranking of the full set of environmental features stays roughly unchanged. Meanwhile, the progenitor galaxy stellar mass rises to third place. Consistently, in \textsc{Magneticum}, $\delta_{1,\mathrm{sp}}$ remains the overwhelmingly dominant predictor, with the progenitor stellar mass rising to second place. Although for galaxies with stellar masses falling within a broader range at $z\sim3$, the stellar mass of their descendant galaxies at $z=0$ would naturally be expected to depend on their stellar mass at $z\sim3$, we find in both simulations that $\delta_{1,\mathrm{sp}}$ has significant impact on stellar mass accumulation. We attribute this to the gravitational collapse of matter within spherical regions, which will be discussed in detail in Section \ref{sec: different-z rfr}.

To further assess the improvement in predictive performance introduced by each environmental metric, we constructed a baseline random forest model with stellar mass as the sole input feature, then added each metric one at a time to this mass-only model to compare the resulting predictions. We computed the logarithmic ratios between the true and predicted stellar masses and defined the scatter as the 20th–80th percentile range of the ratios. In \textsc{Eagle}, the inclusion of $\delta_{1,\mathrm{sp}}$ reduces the scatter by $\sim$0.16 dex, while both $\delta_{4,\mathrm{sp}}$ and $\Sigma_3$ lead to reductions of $\sim$0.15 dex compared to the mass-only model. In \textsc{Magneticum}, adding $\delta_{1,\mathrm{sp}}$ results in a much larger reduction in scatter than $\delta_{4,\mathrm{sp}}$ and $\Sigma_3$, consistent with its significantly higher relative importance. The scatter decreases by about 0.60 dex with $\delta_{1,\mathrm{sp}}$, whereas both $\delta_{4,\mathrm{sp}}$ and $\Sigma_3$ yield reductions of $\sim$0.36 dex. In both simulations, metrics that are less important than the three top-ranked ones also produce smaller reductions in scatter.

Finally, while the lower mass limit ($10^{9}M_{\odot}$) of galaxies included in the calculation of environmental metrics is set to ensure a sufficient number of stellar particles in each galaxy, we also examine whether a higher mass cut would affect the final ranking of these metrics. We find that in \textsc{Eagle}, when only galaxies with stellar masses above $10^{9.5}M_{\odot}$ and $10^{10}M_{\odot}$ are considered, respectively, the ranking of $\delta_{1,\mathrm{sp}}$ drops to the 2nd and 5th places, with the corresponding $\mathrm{R.I.}$ of about $11\%$ and $9\%$. Meanwhile, $\delta_{4,\mathrm{sp}}$ becomes the most important metric in both cases, with $\mathrm{R.I.}$ of $\sim$20$\%$ and $\sim$12$\%$, respectively. In \textsc{Magneticum}, $\delta_{1,\mathrm{sp}}$ remains the most important metric when the calculation is restricted to galaxies with $M_{*}>10^{9.5}M_{\odot}$, though its importance decreases to $\sim$25$\%$. When the mass cut is increased to $10^{10}M_{\odot}$, $\delta_{1,\mathrm{sp}}$ ranks third ($\mathrm{R.I.}\sim10\%$), while $\delta_{4,\mathrm{sp}}$ ranks first ($\mathrm{R.I.}\sim14\%$). The relatively sparse distribution of more massive galaxies makes their local environments on small scales (e.g., 1 cMpc) nearly indistinguishable. A larger aperture (e.g., 4 cMpc) is therefore required to encompass enough neighbouring galaxies such that the fixed-aperture measures can exhibit meaningful variations. Also, given that more massive haloes tend to cluster more strongly \citep[e.g.,][]{1996Mo,2001Sheth,2005Gao,2010Tinker,2011Zehavi}, the presence of a greater number of massive galaxies on larger scales may indicate that the target galaxy resides in a locally high-density environment. However, this connection is indirect and may account for the reduced relative importance of the top-ranked metrics. This reduced importance also suggests that, when considering higher mass cuts, the environment as a whole tends to lose its discriminating power in predicting subsequent mass evolution.

\section{Progenitor and Descendant Matching Considering the Environment}\label{sec:results}

To more specifically analyse the role of the environment in the stellar mass accumulation of galaxies, and to assess whether considering high-redshift environments helps in connecting high-redshift galaxies to their low-redshift descendants, we select galaxies from each simulation with stellar masses in the ranges of $\log (M_{*}/M_{\odot})=9.1, 9.6, 10.1, 10.6\,\pm0.1\,\mathrm{dex}$, and track their number density/stellar mass evolution using merger trees. In \textsc{Eagle}, for the mass bin with a median stellar mass of $\log (M_{*}/M_{\odot})=10.6 $, we set the bin width to $\pm0.2\,\mathrm{dex}$ to ensure a sufficient number of galaxies for statistical analysis. In \textsc{Eagle}, we find 1209, 442, 279, and 117 galaxies in the respective mass bins, whereas in \textsc{Magneticum}, we find 2253, 528, 245 and 100 galaxies in the corresponding mass bins. Based on the results from the RF analysis in Section \ref{sec:random forest results}, we select $\delta_{1,\mathrm{sp}}$, as it is identified as one of the best predictors in both simulations for probing fixed-scale environments. We further subdivide the galaxies in each mass range into three bins based on environmental percentiles, which are determined by $\delta_{1,\mathrm{sp}}$: 0--30th, 30--70th, and 70--100th percentiles. We refer to these environmental density bins as low-, medium-, and high-density regimes, and subsequently track the evolution of galaxies within each regime separately.

\subsection{Number Density Evolution}

\begin{figure}
    \centering
    \includegraphics[width=\linewidth]{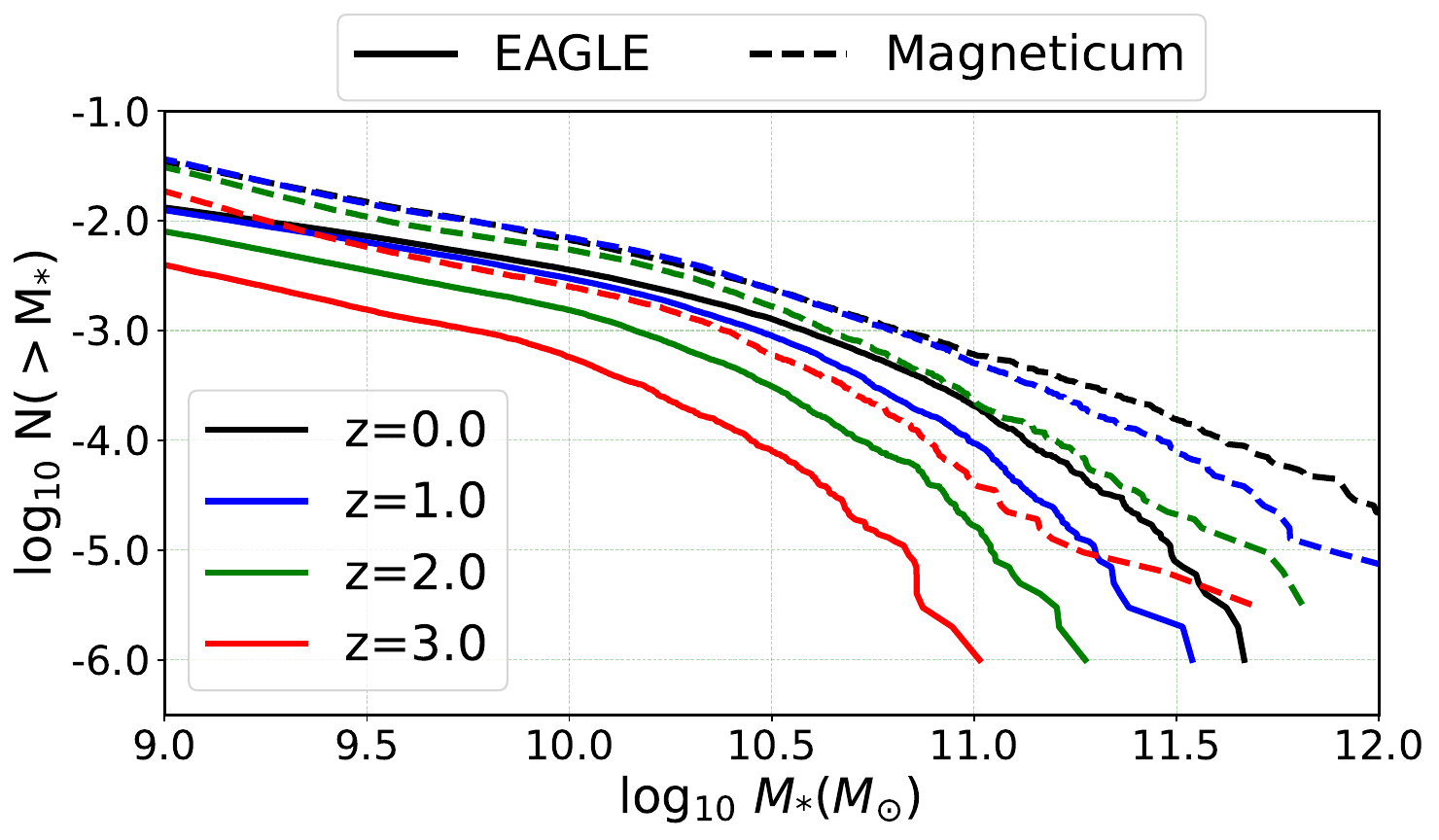}
    \caption{The cumulative mass functions (CMFs) of all galaxies with $\log (M_{*}/M_{\odot})>9.0$ at $z\sim0$ (black), $z\sim1$ (blue), $z\sim2$ (green) and $z\sim3$ (red) in \textsc{Eagle} (solid lines) and \textsc{Magneticum} (dashed lines).}
    \label{fig:CMF}
\end{figure}

To obtain the number density evolution of the target galaxies, we rank the galaxies in each snapshot in the order of decreasing stellar mass and assign each galaxy a rank $R$, where the most massive galaxy is ranked as $R=1$, and so on. A galaxy with rank $R$ will then have a number density $N = \log(R/V)$, where $V$ is the comoving volume of the simulation box. Stellar mass and cumulative number density are therefore interchangeable through the CMF, as they are in a one-to-one mapping relationship. The CMFs of all the galaxies at $0\leq z\leq 3$ with $M_{*}\geq10^{9}M_{\odot}$ in two simulations are shown in Figure \ref{fig:CMF}, where \textsc{Eagle} is plotted with solid lines and \textsc{Magneticum} with dashed lines. Figure \ref{fig:CMF} shows that, due to the larger number of galaxies and the smaller simulation volume, the number density of galaxies at fixed stellar mass is higher in \textsc{Magneticum} than in \textsc{Eagle}. For instance, at $z\sim3$, the number density of galaxies with $9.0 < \log (M_{*}/M_{\odot}) < 10.5$ in \textsc{Magneticum} is typically $\sim0.5$ dex higher than that of galaxies with similar stellar masses in \textsc{Eagle}. The CMF in this mass range also exhibits similar slopes in both simulations; at $z\sim3$, the number density decreases from $-2.40$ to $-4.10\,\rm cMpc^{-3}$ in \textsc{Eagle}, while in \textsc{Magneticum} it decreases from $-1.73$ to $-3.23\,\rm cMpc^{-3}$. In contrast, for galaxies with $\log\,(M_{*}/M_{\odot})>10.5$, the number density decreases more sharply as stellar mass increases in both simulations. 
Additionally, galaxies at similar redshifts in \textsc{Magneticum} exhibit a higher stellar mass limit: at $z\sim3$, the stellar mass of the most massive galaxies in \textsc{Magneticum} reaches $\log (M_{*}/M_{\odot})>11.5$, while in \textsc{Eagle} there are no galaxies with $\log (M_{*}/M_{\odot})>11.0$. These differences likely reflect variations in the implementation of baryonic physics (e.g., AGN feedback)  and the way model parameters are tuned in the two simulations (see Section \ref{sec:method}), which is beyond the scope of this paper. 

\begin{figure}
    \centering
    \includegraphics[width=\linewidth]{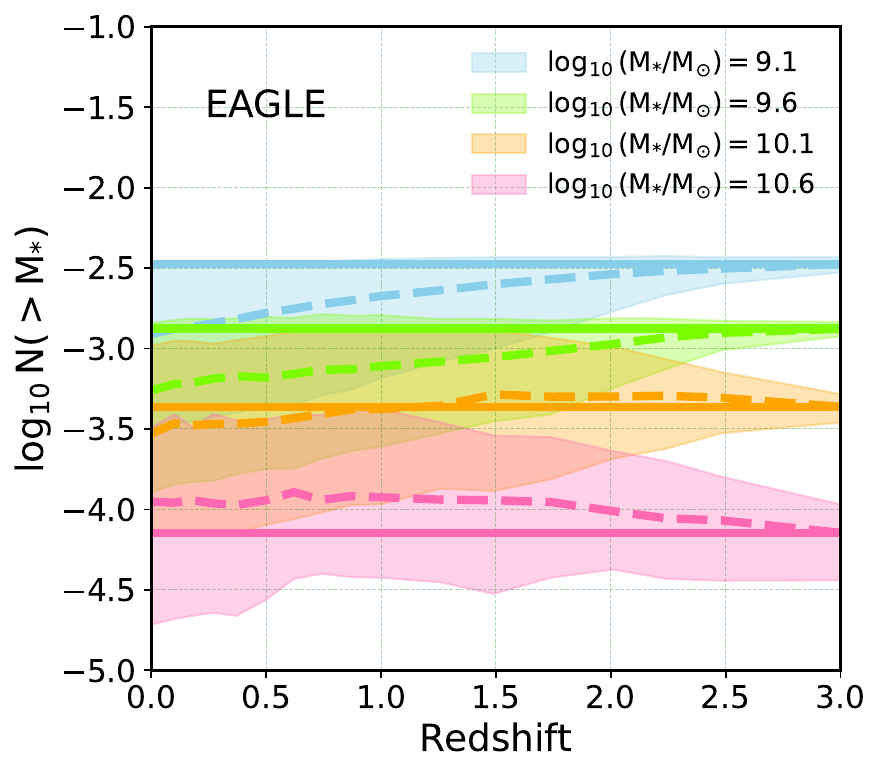}
    \includegraphics[width=\linewidth]{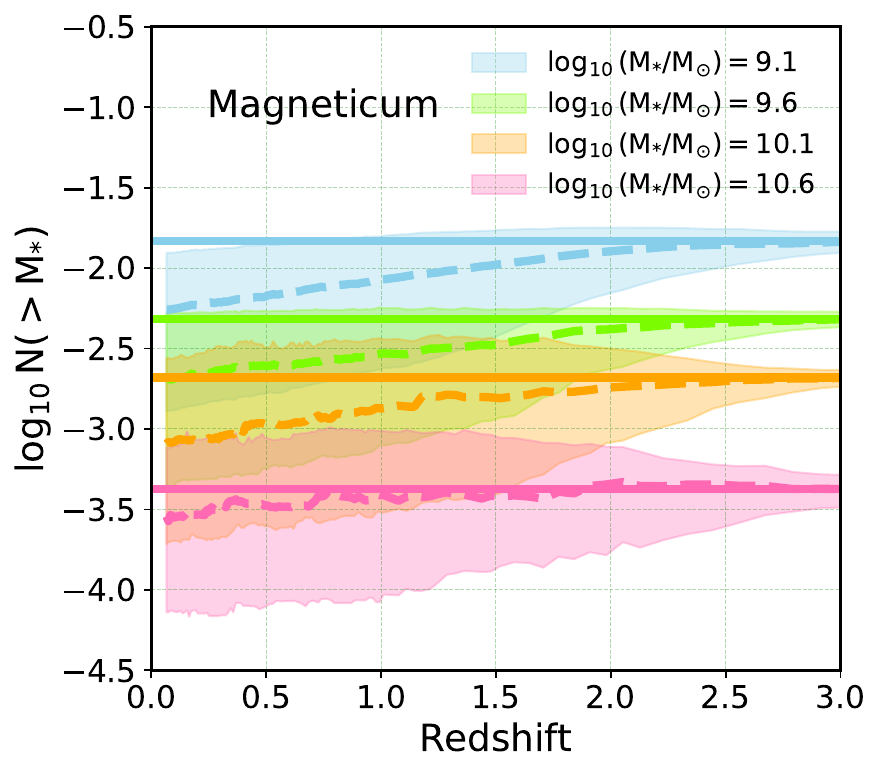}
    \caption{The number density evolution of $z\sim3$ galaxies with $\log (M_{*}/M_{\odot}) \sim 9.1$ (blue), 9.6 (green), 10.1 (yellow), and 10.6 (red) in \textsc{Eagle} (top) and \textsc{Magneticum} (bottom). The horizontal solid lines indicate a constant number density for reference, while the dashed lines show the median number density evolution within each mass bin. The shaded regions represent the scatter between the 20th and 80th percentiles of the number density distribution at each redshift.}
    \label{fig:CND All}
\end{figure}

Previous studies have demonstrated that using stellar velocity dispersion or dark matter halo mass to assign a number density produces an evolution similar to that obtained using stellar mass \citep{2015Torrey}. We first present our analysis of number density evolution, as it is less sensitive to variations in baryonic physics implementations across different simulations compared to stellar mass. In each mass bin, if two or more galaxies merge at a certain snapshot, their shared descendant will only be counted once after this point. The evolution of the number densities for galaxies across four mass bins is shown in Figure \ref{fig:CND All}, with \textsc{Eagle} in the upper panel and \textsc{Magneticum} in the lower panel. The horizontal solid lines represent a constant number density evolution, assuming the entire population with similar stellar masses retains the same median number density as observed at $z\sim3$ throughout its evolution. The dashed lines indicate the median number density evolution within each mass bin, as directly derived from the merger tree. By definition, these two curves coincide at the starting point of the tracking, around $z\sim3$. The shaded regions show the 20th and 80th percentiles of the number density distributions for the populations that were originally in each stellar mass bin. The median number densities and the scatters in number density of the $z\sim0$ descendants of galaxies in different environments are summarised in Table \ref{tab:merger_tree_results}.

\begin{figure*}
    \centering
    \includegraphics[width=\linewidth]{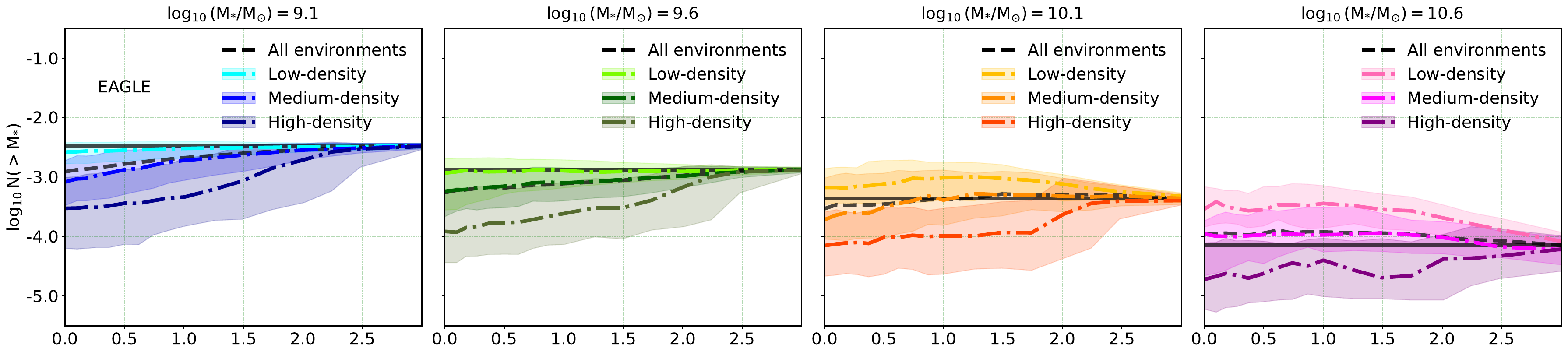}
    \includegraphics[width=\linewidth]{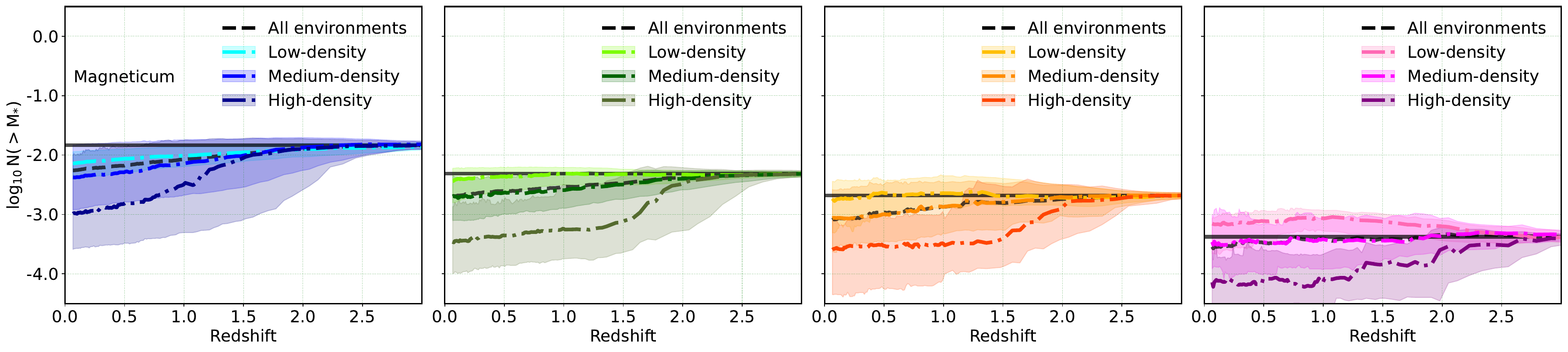}
    \caption{From left to right: the number density evolution of $z\sim3$ galaxies with $\log (M_{*}/M_{\odot}) \sim 9.1$, 9.6, 10.1, and 10.6 in different environments, characterised by the 1-cMpc spherical overdensity($\delta_{1,\mathrm{sp}}$). The top row shows results from the \textsc{Eagle} simulation, while the bottom row corresponds to the \textsc{Magneticum} simulation. In each panel, the solid and dashed black lines have the same meaning as those in Figure \ref{fig:CND All}. The median number density evolution tracks and the scatters between the 20th and 80th percentiles of the number density distribution of galaxies belonging to each environmental density bin are indicated by dotted-dashed lines and shaded areas in different colours, respectively.}
    \label{fig:CND Bins}
\end{figure*}

In \textsc{Eagle}, a notable finding is that, for galaxies with median stellar masses of $\log (M_{*}/M_{\odot})\sim9.1$ and 9.6, the median number density derived from the merger tree starts to diverge from the constant number density from $z\sim3$, with the deviation increasing over time. At $z\sim0$, the differences between the constant and median number densities are 0.44 and 0.38 $\mathrm{dex}$ for the two bins, respectively. This suggests that galaxies in these mass bins show an overall upward shift in their position within the stellar mass ranking of the entire galaxy population in the simulation box. The deviation from a constant number density diminishes with increasing stellar mass. For galaxies with $\log (M_{*}/M_{\odot})\sim10.1$, the median evolution roughly aligns with a constant number density, while for galaxies with $\log (M_{*}/M_{\odot})\sim10.6$, the median number density consistently remains $\sim 0.2\,\mathrm{dex}$ higher than the corresponding constant number density. 
Another significant observation is that, despite the relatively narrow scatter in number density at $z\sim3$ for each bin, the scatter increases substantially by $z\sim0$ when considering the merger tree-based evolution. The number density scatters for the four mass bins at $z\sim0$ are 0.99, 1.05, 1.18, and 1.22 dex, respectively. This highlights the diversity in stellar mass assembly among galaxies, even when they are initially selected from a narrow mass range at $z\sim3$.

In \textsc{Magneticum}, aside from differences in absolute values, the median number density evolution for the three mass bins with $\log (M_{*}/M_{\odot})\sim9.1$, 9.6 and 10.1 consistent lies below the constant number density evolution, with the deviation increasing over time. For these three bins, the differences between the constant and median number densities at $z\sim0$ are 0.42, 0.37, and 0.40 $\,\mathrm{dex}$, respectively. In contrast, for galaxies with $\log (M_{*}/M_{\odot})\sim10.6$, the overall number density evolution closely resembles the constant number density evolution, similar to what is observed for the $\log (M_{*}/M_{\odot})\sim10.1$ bin in \textsc{Eagle}. Similar to \textsc{Eagle}, the merger tree tracking method in \textsc{Magneticum} introduces significant scatter, which increases with decreasing redshift. At $z\sim0$, the 20th to 80th percentile scatters for the number densities in the four mass bins are all $\sim1\,\mathrm{dex}$.

We further subdivide galaxies at $z\sim3$ across different mass ranges into three environmental density bins based on $\delta_{1,\mathrm{sp}}$ to investigate the influence of the external environment at the corresponding scale on number density evolution. The results are shown in Figure \ref{fig:CND Bins}. In each figure, the solid and dashed lines, both shown in black, correspond to the constant and median number density evolution, respectively, of all galaxies within each mass bin, as shown in Figure \ref{fig:CND All}, while the median number density evolution tracks and the scatter between the 20th and 80th percentiles of the number density distribution of galaxies belonging to each environmental density bin are indicated by dotted-dashed lines and shaded areas of different colours. The corresponding values are listed in Table \ref{tab:merger_tree_results}.

In \textsc{Eagle}, the number density evolution shows significant differences across environmental density bins. The median number density evolution of galaxies in medium-density environments is similar to the overall median evolution, whereas galaxies in the low- and high-density environments consistently exhibit higher and lower median number densities, respectively. Note that in the first two mass bins, the median number densities of galaxies in low-density environments stay close to the number densities expected from constant evolution, whereas in the last two mass bins, they remain higher. In contrast, the median number densities in high-density environments consistently fall below the constant number densities across all stellar mass bins.

Moreover, for each mass bin, the scatter in number density at $z\sim0$ tends to be larger for galaxies originating from higher-density environments. For galaxies in the mass bins with $\log(M_{*}/M_{\odot}) = 9.1$, 9.6, 10.1 and 10.6, the number density scatters at $z\sim0$ for the low-, medium- and high-density environments are (0.36, 0.76, 1.19), (0.86, 0.75, 1.16), (0.61, 1.03, 1.07), and (0.68, 0.92, 1.01) dex, respectively. This reflects the diverse pathways of stellar mass accumulation for galaxies in higher-density environments.

Similar to \textsc{Eagle}, galaxies in \textsc{Magneticum} with comparable initial stellar masses at $z\sim3$ exhibit distinct differences in number density evolution depending on their initial external environments. Galaxies that initially reside in higher-density environments consistently show lower median number densities than those in lower-density environments throughout the subsequent evolution. The median number densities of galaxies that were in high-density environments also remain below the constant number densities across all mass bins. Meanwhile, the median number density evolution of galaxies in low-density environments exhibits a mass-dependent relationship with the constant evolution: it lies below the constant track at low stellar mass and exceeds the constant track at high stellar mass. Additionally, in \textsc{Magneticum}, for each mass bin, the scatters in number density of the $z\sim0$ descendants of galaxies in different environments exhibit a similar ranking, with the smallest scatter observed in low-density environmental bins and the largest in high-density environmental bins. For the four mass bins, the scatters at $z\sim0$ for the low-, medium-, and high-density environmental density bins are (0.46, 1.09, 1.58), (0.58, 0.76, 1.13), (0.86, 1.04, 1.25) and (0.60, 0.92, 0.93) dex, respectively.

\subsection{Stellar Mass Evolution}

\begin{figure}
    \centering
    \includegraphics[width=\linewidth]{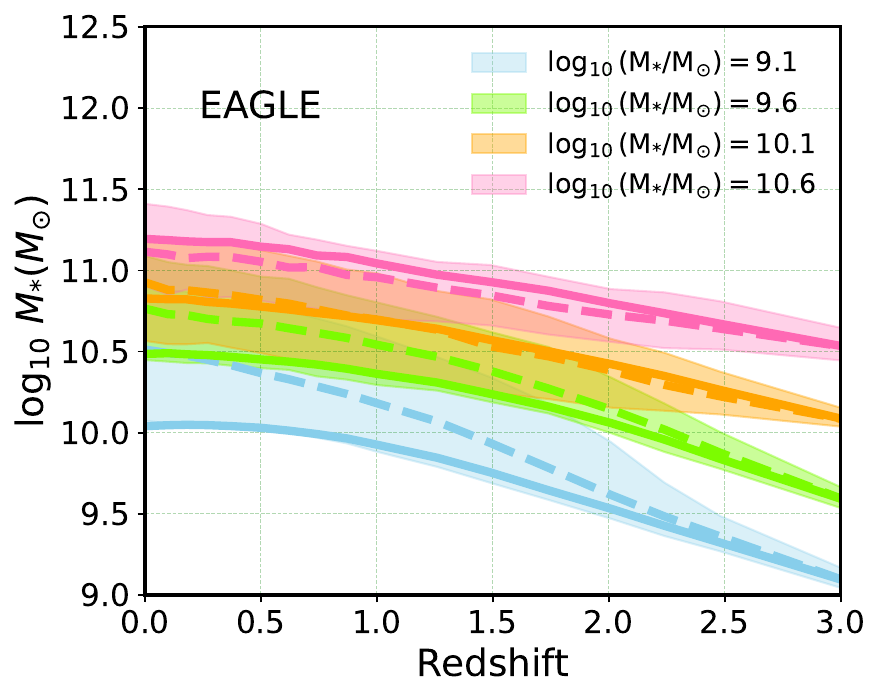}
    \includegraphics[width=\linewidth]{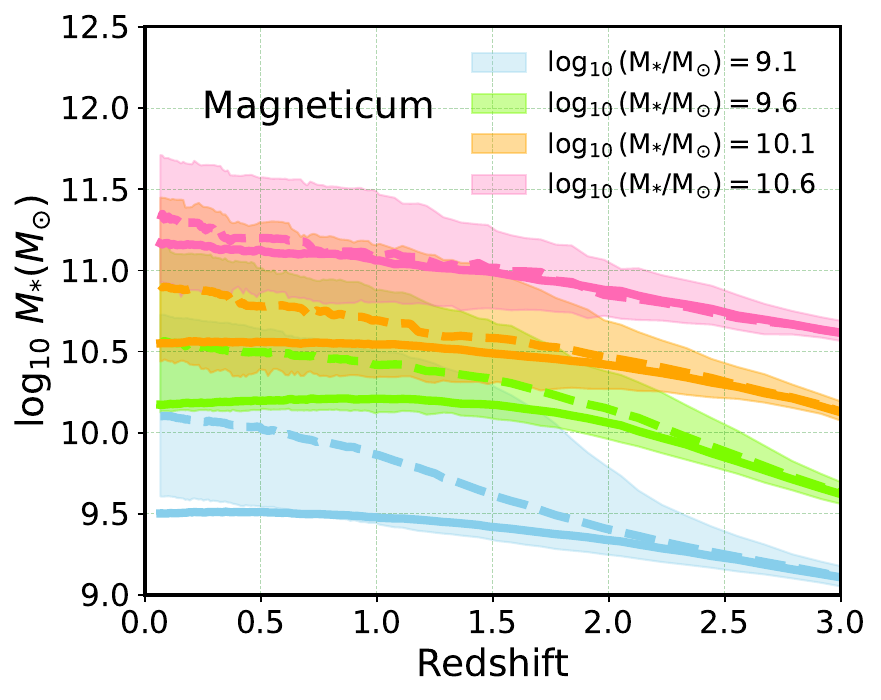}
    \caption{The stellar mass evolution of $z\sim3$ galaxies with $\log (M_{*}/M_{\odot}) \sim 9.1$ (blue), 9.6 (green), 10.1 (yellow), and 10.6 (red) in \textsc{Eagle} (top) and \textsc{Magneticum} (bottom). The solid and dashed lines represent the stellar mass growth tracks corresponding to the constant number density evolution and the median number density evolution, respectively. The shaded regions indicate the range between the 20th and 80th percentiles of the stellar mass distribution at each redshift for the population initially selected at $z\sim3$.}
    \label{fig:Mstar All}
\end{figure}

The number density evolution tracks shown in Figure \ref{fig:CND All} are directly mapped to the stellar mass growth tracks shown in Figure \ref{fig:Mstar All} via the CMFs. In Figure \ref{fig:Mstar All}, solid and dashed lines represent the stellar mass growth tracks corresponding to constant number density evolution and median number density evolution, respectively. The shaded regions indicate the range between the 20th and 80th percentiles of stellar mass distribution at each redshift for the entire population initially selected at $z\sim3$. 

From the figure, it is evident that the stellar mass growth tracks based on constant number density evolution and median number density evolution are offset from each other. For galaxies with $\log(M_{*}/M_{\odot})\sim9.1$ at $z\sim3$, the median stellar mass growth along the merger tree exceeds that inferred from constant number density evolution, resulting in a $\sim0.5$ dex higher stellar mass at $z\sim0$. In the $\log(M_{*}/M_{\odot})\sim9.6$ mass bin, this difference decreases to $\sim0.3$ dex, whereas in the $\log(M_{*}/M_{\odot})\sim10.1$ mass bin, the two approaches converge to similar results. Finally, for galaxies in the $\log(M_{*}/M_{\odot})\sim10.6$ mass bin, the median stellar mass of their descendants at $z\sim0$, as derived from merger trees, is $\sim0.1$ dex lower than that inferred from the constant number density approach. 
In \textsc{Magneticum}, the median stellar masses of descendant galaxies at $z \sim 0$ differ from those inferred from constant number density evolution by approximately 0.60, 0.39, 0.34, and 0.15 dex across the four mass bins, with the discrepancy decreasing progressively toward higher initial mass.

Additionally, as the initial stellar mass at $z\sim3$ increases, the scatter in the stellar mass of descendant galaxies gradually decreases. In \textsc{Eagle}, the stellar mass scatters at $z\sim0$ for the descendants of galaxies initially in the $\log(M_{*}/M_{\odot})\sim9.1$, 9.6, 10.1, and 10.6 mass bins at $z\sim3$ are 0.83, 0.64, 0.64, and 0.51 dex, respectively. In \textsc{Magneticum}, the corresponding values are 1.11, 1.05, 1.01, and 0.79 dex, respectively. This gradual reduction in scatter can be partially explained by the steep decline in the CMFs at the high-mass end, as shown in Figure \ref{fig:CMF}. It is also evident that, in \textsc{Magneticum}, galaxies with similar initial stellar masses have descendants whose stellar mass scatters at $z\sim0$ are generally $\sim0.3$ dex larger than that in \textsc{Eagle}.

\begin{figure*}
    \centering
    \includegraphics[width=\linewidth]{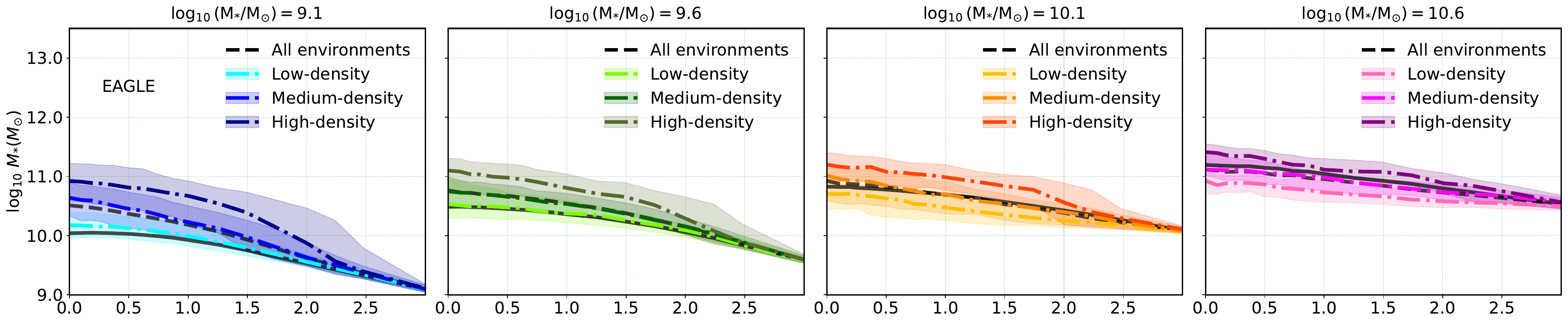}
    \includegraphics[width=\linewidth]{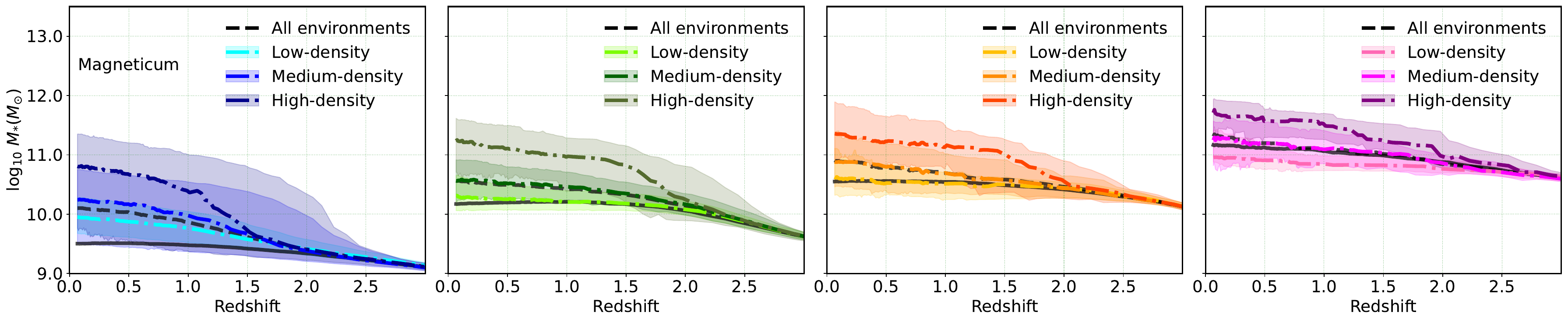}
    \caption{From left to right: the stellar mass evolution of $z\sim3$ galaxies with $\log (M_{*}/M_{\odot}) \sim 9.1$, 9.6, 10.1, and 10.6 in different environments, characterised by the 1-cMpc spherical overdensity($\delta_{1,\mathrm{sp}}$). The top row shows the results from the \textsc{Eagle} simulation, while the bottom row corresponds to the \textsc{Magneticum} simulation. In each panel, the solid and dashed lines have the same meaning as those with the same colour in Figure \ref{fig:Mstar All}. The median stellar mass evolution tracks and the
    scatter between the 20th and 80th percentiles of the stellar mass 
    distribution of galaxies belonging to each environmental density bin
    are indicated by dotted-dashed lines and shaded areas in different colours, respectively.}
    \label{fig:Mstar Bins}
\end{figure*}

The stellar mass growth tracks for galaxies in different environments are shown in Figure \ref{fig:Mstar Bins}. The solid and dashed black lines represent the stellar mass growth tracks for constant number density and the median number density evolution within each mass bin, respectively, as shown in Figure \ref{fig:Mstar All}. The median stellar mass growth tracks and the scatter between the 20th and 80th percentiles of the stellar mass distribution at each redshift are indicated by dotted-dashed lines and shaded areas in different colours for each environmental density bin. All values at $z\sim0$ described above are also summarised in Table \ref{tab:merger_tree_results}.

According to the median growth tracks, galaxies residing in higher-density $z\sim3$ environments tend to accumulate stellar mass more rapidly in both \textsc{Eagle} and \textsc{Magneticum}. Compared to galaxies in medium-density environments, those in high-density environments grow their stellar mass more rapidly, while those in low-density environments experience slower growth. As a consequence, in \textsc{Eagle}, the external environment, as measured by $\delta_{1,\mathrm{sp}}$, leads to differences $\sim$0.5--0.75 dex in the median stellar mass between the $z\sim0$ descendants of galaxies that originally resided in overdense and underdense regions at $z\sim3$. In other words, across the four mass bins, the median stellar mass of the $z\sim0$ descendants of galaxies in $z\sim3$ overdense regions is roughly 3--5 times that of the descendants of galaxies in underdense regions. In comparsion, in \textsc{Magneticum}, those that originated in higher-density environments exhibit a median stellar mass that is $\sim$0.75--0.9 dex (i.e., $\sim$5--8 times) higher than their counterparts from lower-density environments. 

Moreover, within the same environmental density bin, the stellar mass scatters of descendants at $z\sim0$ decrease with increasing initial stellar mass of the progenitor galaxies. This trend holds in both \textsc{Eagle} and \textsc{Magneticum}, suggesting that considering the external environment allows for a more robust connection between massive galaxies at $z\sim3$ and their descendants at $z\sim0$.

Finally, we note that in cases where the stellar mass growth track inferred from constant number density evolution falls below the overall median growth track, it may coincide with the median growth track of galaxies residing in the low-density environments. However, the initial stellar mass bins in which galaxies display this correspondence during their evolution differ between the two simulations.

\begin{table*}
\renewcommand{\arraystretch}{1.2}
    \centering
    \caption{Median and scatter of number density and stellar mass of $z\sim0$ descendants of $z\sim3$ galaxies by initial stellar mass and environment. Stellar mass is expressed on a logarithmic scale, with units in solar masses.}
    \begin{tabular}{c c | cccc | cccc}
        \hline
        \multicolumn{2}{c|}{} & \multicolumn{4}{c|}{Number Density} & \multicolumn{4}{c}{Stellar Mass} \\
        \hline
        Simulation & Initial Mass & All & Low-$\delta_{1,\mathrm{sp}}$ & Medium-$\delta_{1,\mathrm{sp}}$ & High-$\delta_{1,\mathrm{sp}}$ & All & Low-$\delta_{1,\mathrm{sp}}$ & Medium-$\delta_{1,\mathrm{sp}}$ & High-$\delta_{1,\mathrm{sp}}$ \\
        \hline
        \textsc{Eagle} & $9.1\pm0.1$ & $-2.91^{+0.41}_{-0.58}$ & $-2.58^{+0.16}_{-0.20}$ & $-3.08^{+0.36}_{-0.40}$ &  $-3.53^{+0.52}_{-0.67}$ &  $10.51^{+0.39}_{-0.44}$ &  $10.17^{+0.21}_{-0.21}$ &  $10.64^{+0.26}_{-0.31}$ & $10.92^{+0.30}_{-0.34}$ \\
              & $9.6\pm0.1$ & $-3.26^{+0.42}_{-0.63}$ & $-2.93^{+0.24}_{-0.62}$ & $-3.24^{+0.32}_{-0.43}$ &  $-3.91^{+0.63}_{-0.53}$ &  $10.76^{+0.32}_{-0.32}$ &  $10.53^{+0.41}_{-0.23}$ &  $10.75^{+0.24}_{-0.23}$ & $11.10^{+0.21}_{-0.32}$ \\
              & $10.1\pm0.1$ & $-3.53^{+0.55}_{-0.63}$ & $-3.17^{+0.32}_{-0.29}$ & $-3.72^{+0.70}_{-0.33}$ &  $-4.15^{+0.56}_{-0.51}$ &  $10.93^{+0.28}_{-0.36}$ &  $10.71^{+0.18}_{-0.24}$ &  $11.01^{+0.12}_{-0.42}$ &  $11.20^{+0.20}_{-0.24}$ \\
              & $10.6\pm0.2$ & $-3.95^{+0.46}_{-0.76}$ & $-3.54^{+0.38}_{-0.30}$ & $-3.95^{+0.22}_{-0.70}$ &  $-4.72^{+0.61}_{-0.50}$ &  $11.11^{+0.30}_{-0.21}$ &  $10.93^{+0.13}_{-0.23}$ & $11.12^{+0.28}_{-0.09}$ & $11.41^{+0.14}_{-0.24}$\\ 
        \hline
        \textsc{Magneticum} & $9.1\pm0.1$ &  $-2.26^{+0.35}_{-0.63}$ &  $-2.14^{+0.19}_{-0.27}$ &  $-2.37^{+0.54}_{-0.55}$ &  $-2.99^{+0.98}_{-0.60}$ & $10.10^{+0.62}_{-0.49}$ & $9.95^{+0.32}_{-0.27}$ & $10.24^{+0.51}_{-0.74}$ & $10.81^{+0.54}_{-1.04}$ \\
                   & $9.6\pm0.1$ &  $-2.69^{+0.41}_{-0.69}$ &  $-2.44^{+0.21}_{-0.37}$ &  $-2.70^{+0.35}_{-0.41}$ &  $-3.46^{+0.59}_{-0.54}$ & $10.56^{+0.61}_{-0.44}$ & $10.31^{+0.34}_{-0.25}$ & $10.56^{+0.35}_{-0.35}$ & $11.24^{+0.38}_{-0.53}$ \\
                   & $10.1\pm0.1$ &  $-3.09^{+0.53}_{-0.62}$ &  $-2.78^{+0.33}_{-0.53}$ &  $-3.06^{+0.48}_{-0.56}$ &  $-3.59^{+0.49}_{-0.76}$ & $10.91^{+0.54}_{-0.47}$ & $10.63^{+0.48}_{-0.31}$ & $10.88^{+0.50}_{-0.42}$ & $11.36^{+0.53}_{-0.44}$ \\
                   & $10.6\pm0.1$ &  $-3.54^{+0.43}_{-0.60}$ &  $-3.16^{+0.09}_{-0.51}$ &  $-3.46^{+0.47}_{-0.45}$ &  $-4.14^{+0.54}_{-0.39}$ & $11.31^{+0.40}_{-0.39}$ & $10.96^{+0.46}_{-0.08}$ & $11.25^{+0.32}_{-0.43}$ & $11.71^{+0.23}_{-0.35}$ \\
        \hline
    \end{tabular}
    \label{tab:merger_tree_results}
\end{table*}

\subsection{Fit to the Number Density Evolution Including Environmental Measurements}

To further assess the extent to which incorporating environmental information improves the prediction of an individual galaxy’s evolutionary track, we adopt the method from \cite{2015Torrey} to fit the subsequent number density evolution tracks of all galaxies at $z\sim3$ with $\log (M_{*}/M_{\odot})>9.0$ in both \textsc{Eagle} and \textsc{Magneticum}. The functional form adopted in our fitting procedure is given by:

\begin{equation}
    \log N(z)=\sum_{i}\sum_{j=0}^{4}C_{i,j}x_{i}^{j},
\end{equation}

\noindent
where $x_{i}$ are the galaxy properties, $j$ is the polynomial order, and $C_{i,j}=c_{0,i,j}+c_{1,i,j}z+c_{2,i,j}z^{2}$ are the redshift-dependent coefficients. Specifically, we perform linear regressions using stellar mass in logarithmic scale ($x_0$) alone and in combination with $\delta_{1,\mathrm{sp}}$ ($x_1$), and compare the results.

\begin{figure}
    \centering
    \includegraphics[width=\linewidth]{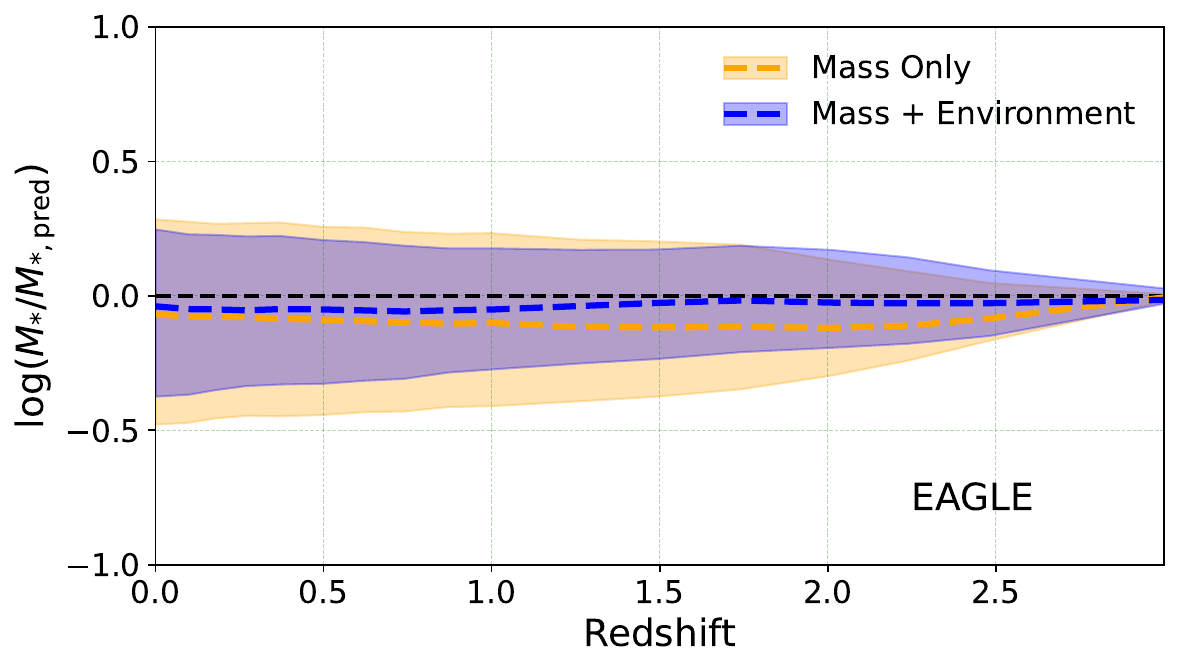}
    \includegraphics[width=\linewidth]{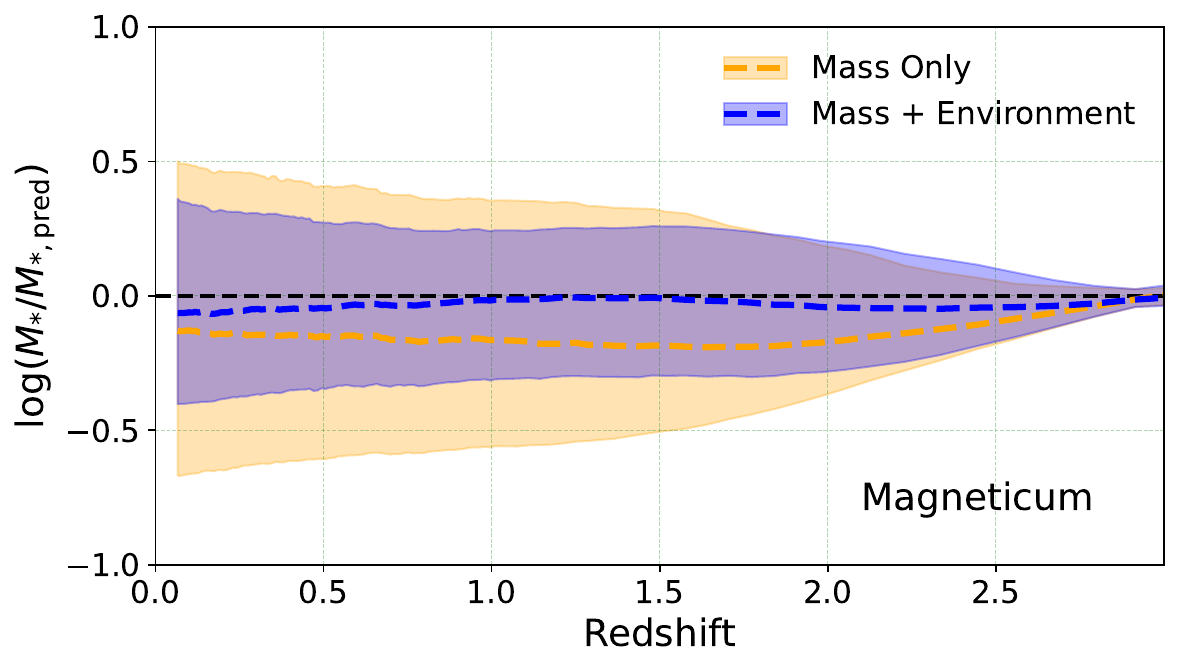}
    \caption{The median and scatter of the logarithmic ratio of the actual to predicted stellar mass as a function of redshift for $z\sim3$ galaxies in \textsc{Eagle}(top) and \textsc{Magneticum}(bottom). Orange and blue dashed lines show the predictions of the median residuals based on stellar mass alone and stellar mass plus environment ($\delta_{1,\mathrm{sp}}$) at $z\sim3$, respectively. Shaded areas of the same colour represent the associated scatter.}
    \label{fig:CND Residual}
\end{figure}

We converted the predicted and actual number density evolution into stellar mass evolution using the CMF shown in Figure \ref{fig:CMF}. Figure \ref{fig:CND Residual} shows the median and the 20–80\% scatter of the logarithmic ratios between the actual and predicted stellar masses over the redshift range $0<z<3$, for \textsc{Eagle} (top panel) and \textsc{Magneticum} (bottom panel). The inclusion of $\delta_{1,\mathrm{sp}}$ results in a median residual closer to 0, along with a reduced scatter in the residuals. At $z\sim0$, the scatter is reduced by $\sim$0.15 dex in \textsc{Eagle}, corresponding to a 20\% decrease, whereas in \textsc{Magneticum}, it is reduced by $\sim$0.4 dex, corresponding to a 35\% decrease. These reductions are comparable to those derived from the random forest analysis in Section \ref{sec:random forest results}, indicating that both approaches produce a similar level of improvement within each simulation. However, the absolute values of the scatter remain large, at $\sim$0.6 dex in \textsc{Eagle} and $\sim$0.75 dex in \textsc{Magneticum}. This scatter is consistent with that obtained from the random forest model described in Section \ref{sec:random forest} that incorporates all the environmental metrics, $\sim$0.66 dex and $\sim$0.85 dex for \textsc{Eagle} and \textsc{Magneticum}, respectively.

Turning to the performance of these regression models in predicting the evolution of $z\sim3$ galaxies in narrow stellar mass bins, we find no significant evolution in either the median residual or the scatter of residuals with varying initial stellar mass in both simulations. We also note that a more physically motivated functional form, rather than a simple polynomial, may improve the predictive power of the external environment as a galaxy parameter in addition to stellar mass. However, even this simple fit already implies a significant improvement over the widely adopted assumption of a constant number density.

\section{Discussion}\label{sec:discussion}
\subsection{Two Growth Pathways: In-situ Star Formation vs Galaxy Merger}
\label{sec:insitu vs merger}

\begin{figure*}
\centering
    \includegraphics[width=\linewidth]{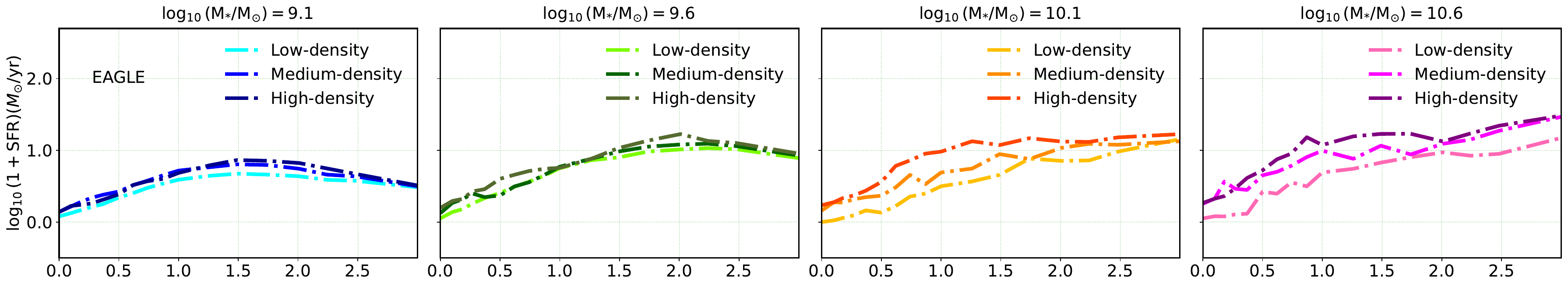}
    \includegraphics[width=\linewidth]{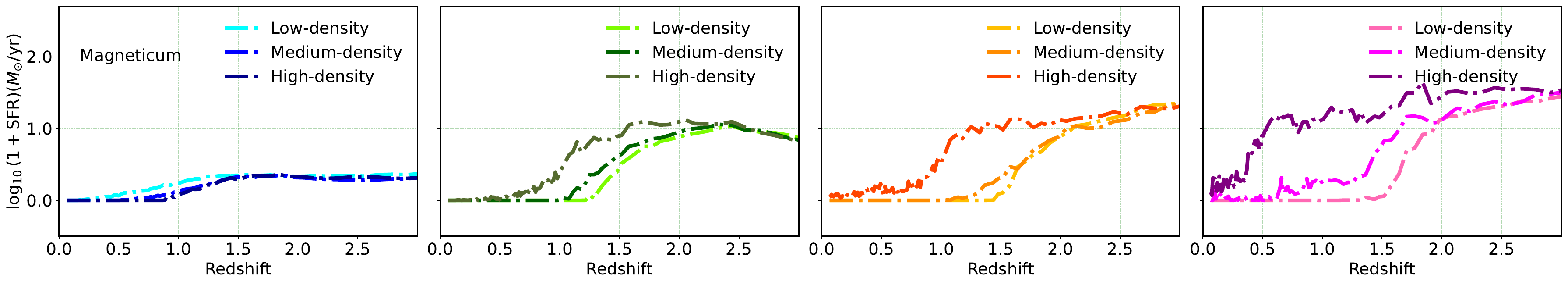}
    \caption{From left to right: the star formation rate evolution of $z\sim3$ galaxies with $\log (M_{*}/M_{\odot}) \sim 9.1$, 9.6, 10.1, and 10.6 in different environments, characterised by the 1-cMpc spherical overdensity($\delta_{1,\mathrm{sp}}$). The top row shows the results from the \textsc{Eagle} simulation, while the bottom row corresponds to the \textsc{Magneticum} simulation. In each panel, the median star formation rate evolution tracks of galaxies belonging to each environmental density bin are indicated by dotted-dashed lines in different colours. For clarity, we omit the median evolution curve of the entire sample in each mass bin.}
    \label{fig:SFR Bins}
\end{figure*}

To further examine how the external environment mentioned above has led to the divergent evolutionary pathways of galaxies with similar initial stellar masses at $z\sim3$, as discussed in Section \ref{sec:results}, we analyse their SFR evolution over time, as well as the amount of external stellar mass they have accreted since $z\sim3$. The median star formation evolution of galaxies in different environments is presented in Figure \ref{fig:SFR Bins}. In both simulations, the median SFR for galaxies with higher initial stellar masses (i.e., $\log(M_{*}/M_{\odot})=10.1$ and $10.6$) steadily decreases since $z\sim3$. In contrast, the quenching process for lower-mass galaxies (i.e., $\log(M_{*}/M_{\odot})=9.1$ and $9.6$) is more gradual, with the median SFR first increasing, peaking between $1<z<3$, and then declining.

In both simulations, galaxies with $\log(M_{*}/M_{\odot})=9.1$ at $z\sim3$ in different environments exhibit similar median SFR. In contrast, for the population with $\log (M_{*}/M_{\odot})\sim10.1$ and $10.6$ at $z\sim3$ in both simulations, galaxies in medium- and high-density environments exhibit enhanced median star formation rates compared to those in low-density environments during subsequent evolution. While it is widely known that AGN feedback tends to be more effective in suppressing in-situ star formation in high-mass halos \citep[e.g.,][]{2017Bower}, the denser 1-cMpc scale environment is likely associated with a larger scale reservoir of gas, which can fuel in-situ star formation. A recent study based on \textsc{Magneticum} also suggests that the mass of surrounding gas correlates with the in-situ stellar mass that galaxies eventually form \citep{2025Remus}. Additionally, it may facilitate a higher frequency of galaxy mergers, which can trigger bursts of star formation. This environment-driven enhancement is also observed in the population with an initial stellar mass of $\log (M_{*}/M_{\odot})\sim9.6$ in \textsc{Magneticum}.

We evaluate the external contribution of galaxy mergers to the stellar mass growth of galaxies across different environments. To achieve this, we first identify all unique $z\sim0$ descendants of galaxies with $\log(M_{*}/M_{\odot}) \sim 9.1$, 9.6, 10.1, and 10.6 at $z\sim3$ using merger trees. Next, we track the evolution of these $z\sim0$ galaxies backward in time along the merger trees until $z\sim3$ and identify all progenitors with $\log(M_{*}/M_{\odot}) > 9.0$. In \textsc{Eagle}, the main branch is defined as the branch with the maximum 'branch mass', i.e., the total mass of all particle species summed over all progenitors on the same branch \citep{2017Qu}. In \textsc{Magneticum}, it is identified using \textsc{L-BaseTree} \citep{2005Springel}. The ex-situ fraction is defined as the sum of the stellar masses of galaxies belonging to the sub-branches divided by the stellar mass of the descendant galaxy at $z\sim0$. During the merging of two galaxies, part of the stellar components begins to overlap and is assigned to the more massive galaxy by the halo-finder. Thus, to calculate the ex-situ fraction, we use the maximum stellar mass of each sub-branch prior to merging. It has been shown that the choice of aperture-based versus \textsc{Subfind}-based stellar mass has essentially no impact on the resulting ex-situ fraction (\citealp[e.g.][]{2017Qu}). By contrast, greater uncertainty may stem from using merger trees rather than explicit particle tracing. A more accurate estimation of the ex-situ fraction could be achieved by tracing particle data across snapshots, but this is beyond the scope of this paper.

\begin{figure}
    \centering
    \includegraphics[width=\linewidth]{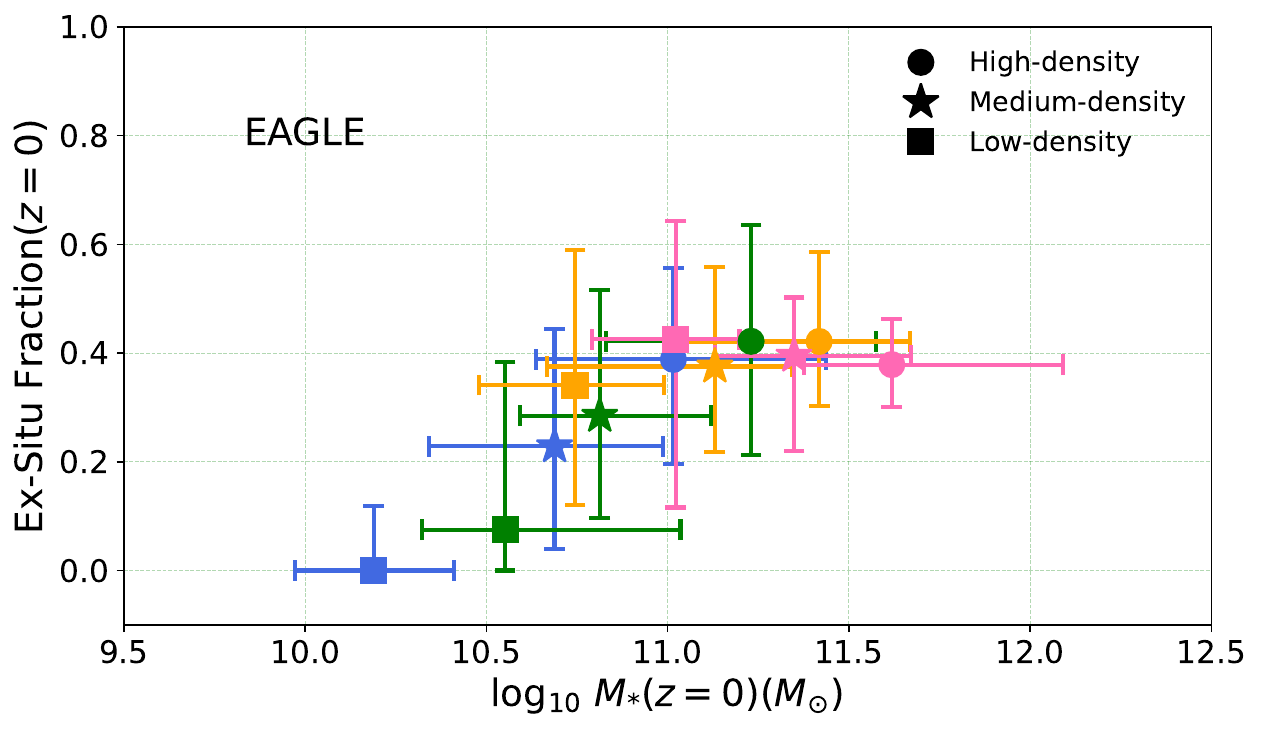}
    \includegraphics[width=\linewidth]{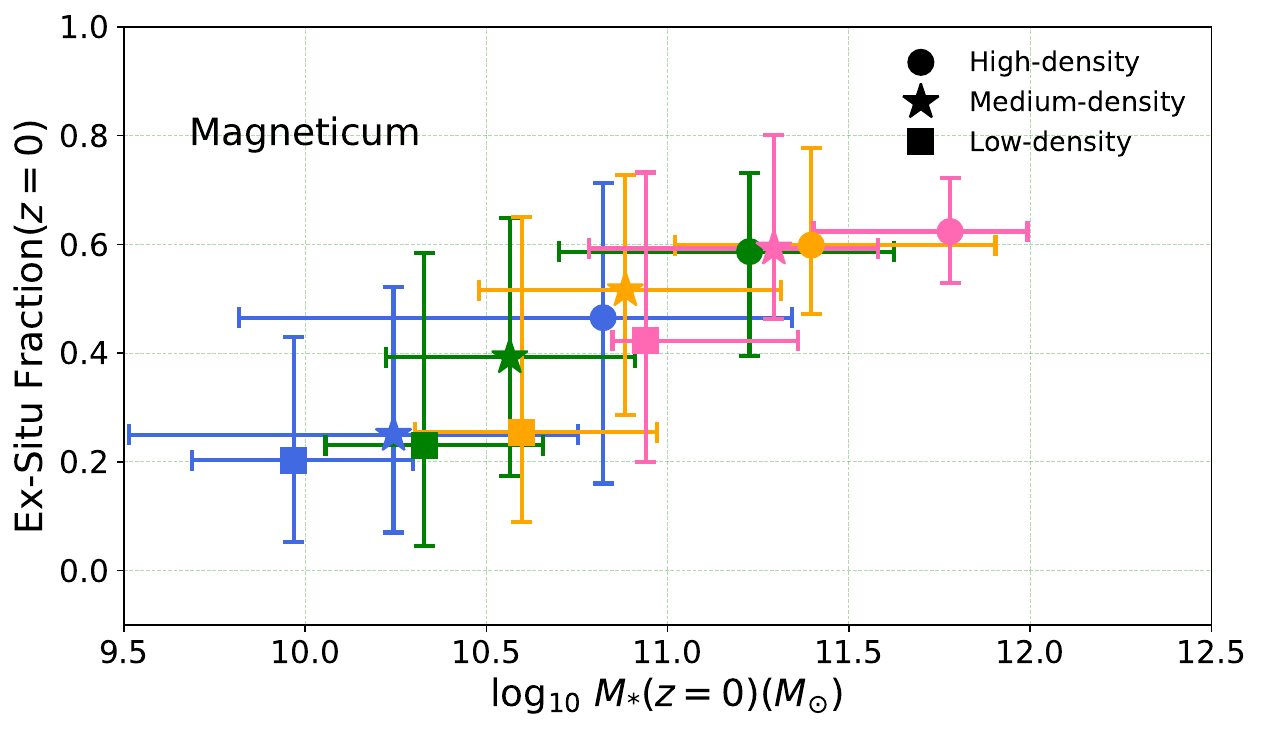}
    \caption{The ex-situ fraction of the $z\sim0$ descendants of $z\sim3$ galaxies with $\log (M_{*}/M_{\odot}) \sim 9.1$ (blue), 9.6 (green), 10.1 (yellow), and 10.6 (red) in \textsc{Eagle} (top) and \textsc{Magneticum} (bottom) as a function of their stellar mass. Circles, stars, and squares represent galaxies from high-, medium-, and low-density environments at $z\sim3$, respectively, as defined in Section \ref{sec:results}.}
    \label{fig:ex-situ fraction}
\end{figure}

In Figure \ref{fig:ex-situ fraction}, we present the ex-situ fraction of the $z\sim0$ descendants of the selected galaxies as a function of their stellar mass. We use the same colours as in the previous figures to distinguish descendant galaxies originating from different initial stellar mass bins. Galaxies from high-, medium-, and low-density environments at $z\sim3$ are represented by circles, stars, and squares, respectively. The error bars represent the range corresponding to the 20th to 80th percentiles of the stellar mass and ex-situ fraction within the galaxy population represented by each data point.

Clearly, in both simulations, the descendants of galaxies with similar initial stellar mass exhibit an overall increasing ex-situ fraction with increasing environmental density at $z\sim3$, from low- to high-density environments. Moreover, the ex-situ fraction differences between galaxies originating from different environments gradually diminish with increasing initial stellar mass. For galaxies with $\log(M_*/M_{\odot})\sim10.6$ that originally resided in medium- and high-density environments at $z\sim3$, the median ex-situ fractions of their $z\sim0$ descendants are very similar in both simulations.

Based on this, we conclude that the different stellar mass growth pathways of galaxies across various environments within a 1-cMpc radius spherical region, as shown in Figure \ref{fig:Mstar Bins}, result from the environmental influence on galaxies of different stellar masses. At the low-mass end (e.g., $\log(M_{*}/M_{\odot})\sim9.1$ and 9.6), the descendants of galaxies that reside in low-density environments exhibit low median ex-situ fractions, indicating that their mass growth primarily relies on in-situ star formation. Notably, a zero ex-situ fraction may result from the mass threshold applied when tracing progenitor galaxies, as we ensure each subhalo contains at least $\sim500$ stellar particles. This finding is consistent with the literature: for galaxies with $\log(M_{*}/M_{\odot}) \lesssim 10.5$ at $z\sim0$, stellar mass growth is primarily driven by in-situ star formation rather than ex-situ accretion \citep{2016Rodriguez-Gomez,2016Dubois,2017Qu,2018Pillepich,2019Tacchella,2020Davison,2022Remus}. Meanwhile, in high-density environments, some low-mass galaxies merge with other galaxies, thus their descendants have higher stellar masses and enhanced ex-situ fractions. The large scatter in both the stellar masses and the ex-situ fractions of their descendants further highlights the diverse evolutionary pathways of low-mass galaxies in the top $30\%$ densest environments.

In contrast, at the high-mass end (e.g., $\log(M_{*}/M_{\odot})\sim10.1$ and 10.6), the stellar mass difference among the descendants of galaxies in different environments is driven by a combination of enhanced in-situ star formation throughout their evolution and greater external stellar mass accretion. The enhanced in-situ star formation may result from increased gas accretion and/or gas-rich mergers. We also note that, for galaxies with the same initial stellar mass, the median ex-situ fraction of their descendants in \textsc{Magneticum} is overall about $10-20\%$ higher than that in \textsc{Eagle}, which is consistent with previous studies (\citealp[e.g.,][]{2022Remus}). Remarkably, even galaxies in the bottom $30\%$ least dense environments can have descendants with ex-situ fractions comparable to those of the descendants of galaxies in the top $30\%$ densest environments. This suggests that $\delta_{1,\mathrm{sp}}$ may not be an optimal proxy for the external stellar mass available for galaxy mass growth during their evolution from $z\sim3$ to $z\sim0$.

\subsection{The Importance of Overdensity Within Different Radii Across Various Redshifts}
\label{sec: different-z rfr}

\begin{figure*}
\centering
    \includegraphics[width=\linewidth]{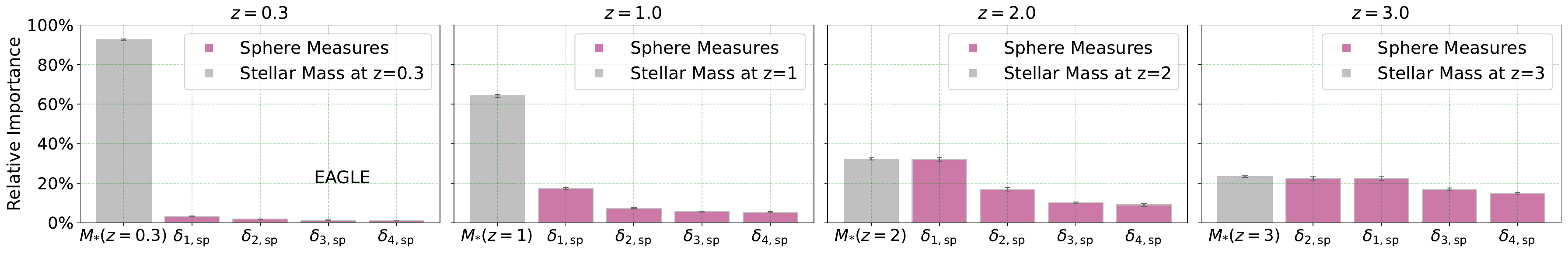}
    \includegraphics[width=\linewidth]{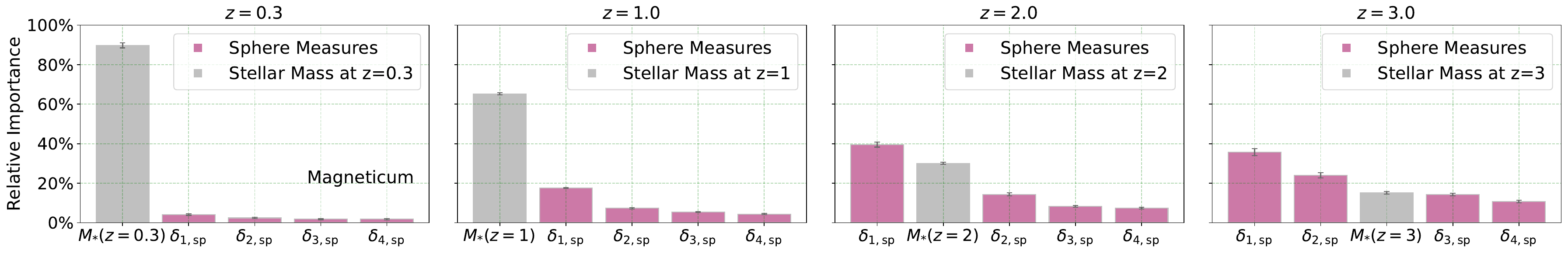}
    \caption{From left to right: the relative importance (R.I.) of the overdensities within spherical regions of different radii and the stellar mass of the progneitor galaxies, in determining the stellar mass of the $z\sim0$ descendants of $z\sim0.3$, 1.0, 2.0, and 3.0 galaxies with $\log (M_{*}/M_{\odot})>9.0$ in \textsc{Eagle} (top) and \textsc{Magneticum} (bottom). The height of each bar denotes the R.I. of each feature, with the error bars representing the standard deviation from 10 independent runs. The bars are colour-coded by feature categories: sphere measures (pink), and the stellar mass of progenitor galaxies (gray).}
    \label{fig:RFR different-z}
\end{figure*}

To investigate whether the stellar mass growth dependence of $z\sim3$ galaxies on $\delta_{1,\mathrm{sp}}$ primarily reflects the gravitational assembly of halos, rather than the baryonic physics implemented by each simulation, we examine how spherical overdensities measured at different radii and redshifts affect the subsequent stellar mass growth of galaxies. In addition to $z\sim3$, we also select galaxies with $\log(M_{*}/M_{\odot})>9.0$ at $z\sim2, 1,$ and 0.3, and identify their $z\sim0$ descendants via merger trees. Previous simulations have shown that the environment plays an important role in driving rapid changes in galaxy morphology, angular momentum, and star-formation activity at $z\sim0.3$, when the Universe was in its middle age \citep{2021Foster}. We calculate the overdensities within spherical regions of $r = 1, 2, 3, 4$ cMpc around each galaxy. These spherical overdensities, together with the initial stellar masses of the galaxies at each redshift, are used as input parameters for the RF analyses to predict the stellar mass of their $z\sim0$ descendants. The R.I. of each feature in determining the descendants’ stellar mass is shown in Figure \ref{fig:RFR different-z}.

Figure \ref{fig:RFR different-z} shows that in both simulations, as the time available for gravitational collapse becomes shorter, the R.I. of the initial stellar mass gradually increases from $\sim15-20\%$ at $z\sim3$ to $\sim90\%$ at $z\sim0.3$. In contrast, while the cosmic time between $z=0$ and $z\sim0.3$ might be too short for the material within the 1-cMpc spherical region around galaxies to have any significant influence on the stellar mass growth of $z\sim0.3$ galaxies, the R.I. of the $\delta_{1,\mathrm{sp}}$ increases from $\sim3\%$ at $z\sim0.3$ to $\sim32\%$ in \textsc{Eagle} and $\sim40\%$ in \textsc{Magneticum} at $z\sim2$, as the available cosmic time until $z\sim0$ doubles. From $z\sim2$ to $z\sim3$, the R.I. of $\delta_{1,\mathrm{sp}}$ decreases in both simulations. Meanwhile, the R.I. of the 2-cMpc spherical region also increases from $\sim2\%$ at $z\sim0.3$ to $\sim20\%$ at $z\sim3$ in both simulations. 

Based on data from both simulations, the RF analyses yield highly similar results, suggesting that the influence of spherical overdensity on stellar mass assembly in both simulations is primarily driven by universal processes such as gravitational collapse, rather than being strongly affected by the different baryonic physics models implemented in each simulation. Even so, the effect of overdensity at the same radius still varies slightly between the two simulations, especially at $z\sim3$. The more precise scale of the external environment that has the greatest impact at different redshifts also remains to be explored in future work. 

More importantly, our RF analyses indicate that even at $z\sim1$, a galaxy's mass can still largely determine the stellar mass of its descendant at $z\sim0$. This suggests that progenitor-descendant matching methods based on stellar mass or number density (\citealp[e.g.,][]{2015Torrey, 2017Torrey}) can still reliably connect galaxies at $z\sim0$ and $z\sim1$(\citealp[e.g.,][]{2018Bezanson,2020Cole}). However, if we aim to study the continuous evolution of galaxies tracing back to higher redshifts (\citealp[e.g., $z\sim2$ or $z\sim3$;][]{2011Brammer,2013Patel,2013vanDokkum,2014Marchesini,2020Stockmann,2020Mendel}), considering the external environments as additional evidence may help improve the accuracy of the matching, as subsequent assembly processes directly related to the external environments can have a significant impact on the stellar mass evolution tracks of galaxies (\citealp[e.g.,][]{2025Kimmig}).

Additionally, if we attempt to connect $z\sim3$ galaxies with their $z\sim0$ descendants, the accuracy of the matching may be improved by first identifying their $z\sim2$ descendants and incorporating the $z\sim2$ environments as additional constraints in the remaining matching process. However, considering the low relative importance of these metrics at $z\lesssim1$, further dividing galaxies with similar stellar masses into subgroups based on their environments at these lower redshifts and tracing their evolution separately is unlikely to significantly enhance the matching precision.

\subsection{The Role of Low-n Nearest neighbour Density}

To determine whether $\Sigma_{3}$ can improve progenitor–descendant matching, we investigate its correlation with $\delta_{1,\mathrm{sp}}$ at $z\sim3$. Both simulations predict through RF analysis that $\Sigma_{3}$ also plays a relatively significant role in the stellar mass growth of $z \sim 3$ galaxies. Since $\Sigma_{3}$ describes the local environment and $\delta_{1,\mathrm{sp}}$ characterises the global environment, understanding their relationship can help build on the discriminative power already provided by $\delta_{1,\mathrm{sp}}$. 

In both simulations, a correlation emerges between the $\delta_{1,\mathrm{sp}}$ and the $\Sigma_{3}$. For galaxies at $z\sim3$ with stellar masses $\log(M_{*}/M_{\odot})>9.0$, the Pearson correlation coefficient between $\delta_{1,\mathrm{sp}}$ and $\Sigma_{3}$ is 0.65 in \textsc{Eagle} and 0.64 in \textsc{Magneticum}. This indicates that galaxies in higher-density global environments tend to also reside in higher-density local environments. Further dividing galaxies into different stellar mass bins, we find that the correlation between $\delta_{1,\mathrm{sp}}$ and $\Sigma_{3}$ tends to strengthen with increasing stellar mass. Specifically, for galaxies with $9.0\leq\log(M_{*}/M_{\odot})9.5$, $9.5\leq\log(M_{*}/M_{\odot})10.0$ and $\log(M_{*}/M_{\odot})\geq10.0$, the coefficients in \textsc{Eagle} are 0.60, 0.63, and 0.75, respectively, while in \textsc{Magneticum} they are 0.62, 0.63, and 0.67.

We further examine the correlation between $\Sigma_{3}$ and the stellar mass of $z\sim0$ descendant galaxies, while controlling for the influence of $\delta_{1,\mathrm{sp}}$. In both simulations, the partial correlation coefficients remain $\lesssim0.2$, suggesting that $\Sigma_{3}$ itself plays a relatively minor role in shaping the stellar mass growth of galaxies. However, its correlation with $\delta_{1,\mathrm{sp}}$ implies that it can still serve as a useful proxy, particularly in observational studies where $\Sigma_{3}$ is more feasible to measure.

\section{Conclusions}\label{sec:conclusion}

Disentangling the influence of physical mechanisms with different effective timescales in galaxy evolution is challenging, and requires accurately tracing the mass growth of galaxies, i.e., matching progenitor and descendant galaxies across cosmic time. In this study we investigate how the external environment of high-redshift galaxies influences their subsequent stellar mass growth tracks. Using galaxies with $\log(M_{*}/M_{\odot}) > 9.0$ at $z \sim 3$ from the \textsc{Eagle} and \textsc{Magneticum} cosmological simulations, we apply random forest regression to assess which environmental metrics best predict the stellar mass of their $z \sim 0$ descendants. Building on the results of the random forest analysis, we examine whether considering the external environment can help improve the accuracy of matching high-redshift galaxies with their descendants in the local Universe in observational studies. Our main results can be summarised as follows:

\begin{itemize}
    \item In both simulations, among all the environmental metrics we consider, the spherical overdensity within a radius of 1.0 cMpc ($\delta_{1,\mathrm{sp}}$) is the best predictor of the subsequent stellar mass assembly of galaxies across different mass ranges at $z\sim3$.
    \item Using the merger tree data, we analyse the subsequent stellar mass growth tracks of galaxies with similar initial stellar mass but residing in different environments, as characterised by $\delta_{1,\mathrm{sp}}$. In both simulations, galaxies with similar stellar mass at $z\sim3$ but different $\delta_{1,\mathrm{sp}}$ values follow markedly different mass growth tracks. Specifically, the median stellar mass of $z\sim0$ descendant galaxies originating from overdense regions is approximately 3--5 times larger than that of their counterparts from underdense regions in \textsc{Eagle}, and 5--8 times larger in \textsc{Magneticum}.
    \item By fitting the number density evolution using a polynomial, we find that incorporating $\delta_{1,\mathrm{sp}}$ as an additional input leads to a more accurate prediction of the median stellar mass evolution and a smaller scatter of the residuals between the predicted and actual stellar masses, compared to using stellar mass at $z\sim3$ as the sole input. In \textsc{Eagle}, the residual scatter at $z\sim0$ is reduced by $\sim$20\%, and in \textsc{Magneticum}, by $\sim$35\%, when environmental information is included.
    
    \item For low-mass galaxies in low-density regions, stellar mass growth is dominated by in-situ star formation. In contrast, low-mass galaxies in high-density environments can remain isolated or undergo mergers, with the latter contributing significantly to stellar mass growth. On the other hand, high-mass galaxies in high-density regions experience both enhanced in-situ star formation--potentially due to a larger available gas reservoir and/or more frequent gas-rich mergers--and increased stellar mass accretion from mergers.
    \item We find that the stellar mass of galaxies at $z \lesssim 1$ remains a strong predictor of the stellar mass of their $z\sim0$ descendants. Thus, number density matching remains effective in linking progenitor and descendant galaxies. However, at $z\gtrsim2$, as the influence of external environmental factors becomes comparable to that of the progenitor's stellar mass, incorporating them as additional constraints may help improve the accuracy of progenitor-descendant matching.
    
\end{itemize} 
We note that, although this study demonstrates that the environment within spherical regions around galaxies can provide additional clues to improve the accuracy of progenitor-descendant matching, a universally applicable correction to galaxy mass growth has yet to be explored. The general trend that galaxies in denser environments tend to grow more in stellar mass stems from the underlying hierarchical framework of galaxy formation, which is shared among the simulations considered here. However, the exact amount of stellar mass growth depends on the detailed implementations of baryonic physics in each simulation, such as the enhancement or suppression of star formation illustrated in Figure \ref{fig:SFR Bins}. Additionally, in practice, accurately determining the 3D positions of galaxies remains challenging due to their peculiar motions. Therefore, it is necessary to explore approximations that can be applied to real observational studies, such as projected environmental tracers (\citealp[e.g.][]{2007Elbaz,2008Cooper,2012Quadri,2017Fossati,2024Jin,2025McConachie,2025Kawinwanichakij}). 

While this study focuses primarily on the evolution of galaxies' global properties (e.g., $M_{*}$, SFR), our results suggest that descendants of high-redshift galaxies in different environments likely differ in their ex-situ stellar mass fraction. Thus, future studies using particle data from cosmological simulations to study the spatially resolved properties (e.g., stellar population profile) of galaxies could offer new insights for progenitor-descendant matching, as they may help distinguish between galaxies of similar stellar mass that formed at different epochs or underwent multiple rejuvenation episodes \citep{2025Fortune}. 

Finally, our study highlights that comprehensive measurements of environments at high redshift are crucial for understanding galaxy evolution across cosmic time. Our results align with findings for protocluster regions \citep{2023Remus}, where the satellite number is a much better predictor of overall future mass growth than the stellar mass of the brightest cluster galaxy. Wide-field slitless spectroscopic observations such as those enabled by \emph{JWST}/NIRISS \citep{2022Willott} or the forthcoming \emph{Roman}/WFI are particularly useful for this purpose, as they not only cover large sky areas and thus encompass a wide range of environments, but also provide more accurate redshifts than photometric methods, thereby enabling more reliable measurements of environments.

\section*{Acknowledgements}
We would like to thank the anonymous referee for providing constructive comments, which have greatly improved the quality and clarity of this manuscript. TG thanks Katherine E. Harborne, Katy Proctor, and Esteban Jimenez Henriquez for helpful discussions. TG, JTM, and CdPL acknowledge support from the Australian Research Council through Discovery Project DP210101945, funded by the Australian Government. LCK acknowledges support by the Deutsche Forschungsgemeinschaft (DFG, German Research Foundation) under project nr. 516355818, the COMPLEX project from the European Research Council (ERC) under the European Union’s Horizon 2020 research and innovation program grant agreement ERC-2019-AdG 882679 and by DFG under Germany's Excellence Strategy -- EXC-2096 -- 3900783311. KG is supported by the Australian Research Council through the Discovery Early Career Researcher Award (DECRA) Fellowship (project number DE220100766) funded by the Australian Government. We acknowledge the Virgo Consortium for making the \textsc{Eagle} simulation data available. The \textsc{Magneticum} simulations were performed at the Leibniz-Rechenzentrum with CPU time assigned to the Project pr83li.

\section*{Data Availability}

The \textsc{Eagle} simulations are publicly available; see \cite{2016McAlpine} and \cite{2017EAGLE} for how to access \textsc{Eagle} data. \textsc{Magneticum} data are partially available at \url{https://c2papcosmosim.uc.lrz.de/} \citep{2017Ragagnin}, with larger data sets on request.



\bibliographystyle{mnras}
\bibliography{example} 







\bsp	
\label{lastpage}
\end{document}